\documentclass[12pt,a4paper]{article}
\usepackage{ifthen} % for conditional statements
\newboolean{pdflatex}
\setboolean{pdflatex}{true} % False for eps figures 

\newboolean{articletitles}
\setboolean{articletitles}{true} % False removes titles in references

\newboolean{uprightparticles}
\setboolean{uprightparticles}{false} %True for upright particle symbols

\usepackage{microtype}
\usepackage{lineno}  % for line numbering during review
\usepackage{xspace} % To avoid problems with missing or double spaces after
\usepackage{graphicx}  % to include figures (can also use other packages)
\usepackage{color}
\usepackage{colortbl}
\graphicspath{{./figs/}} % Make Latex search fig subdir for figures

\usepackage{amsmath} % Adds a large collection of math symbols
\usepackage{amssymb}
\usepackage{amsfonts}
\usepackage{upgreek} % Adds in support for greek letters in roman typeset

\newcommand*\patchAmsMathEnvironmentForLineno[1]{%
\expandafter\let\csname old#1\expandafter\endcsname\csname #1\endcsname
\expandafter\let\csname oldend#1\expandafter\endcsname\csname
end#1\endcsname
 \renewenvironment{#1}%
   {\linenomath\csname old#1\endcsname}%
   {\csname oldend#1\endcsname\endlinenomath}%
}
\newcommand*\patchBothAmsMathEnvironmentsForLineno[1]{%
  \patchAmsMathEnvironmentForLineno{#1}%
  \patchAmsMathEnvironmentForLineno{#1*}%
}
\AtBeginDocument{%
\patchBothAmsMathEnvironmentsForLineno{equation}%
\patchBothAmsMathEnvironmentsForLineno{align}%
\patchBothAmsMathEnvironmentsForLineno{flalign}%
\patchBothAmsMathEnvironmentsForLineno{alignat}%
\patchBothAmsMathEnvironmentsForLineno{gather}%
\patchBothAmsMathEnvironmentsForLineno{multline}%
}

\usepackage{hyperref}    % Hyperlinks in references
\usepackage[all]{hypcap} % Internal hyperlinks to floats.

\def\lhcb {LHCb\xspace}
\def\ux85 {UX85\xspace}

\ifthenelse{\boolean{uprightparticles}}%
{

 \def\Ppi         {\ensuremath{\uppi}\xspace}

 \def\PDelta      {\ensuremath{\Delta}\xspace}                 
 \def\PXi      {\ensuremath{\Xi}\xspace}                 
 \def\PLambda      {\ensuremath{\Lambda}\xspace}                 
 \def\PSigma      {\ensuremath{\Sigma}\xspace}                 
 \def\POmega      {\ensuremath{\Omega}\xspace}                 
 \def\PUpsilon      {\ensuremath{\Upsilon}\xspace}                 
 
 \def\PB      {\ensuremath{\mathrm{B}}\xspace}                 
                  
 \def\PD      {\ensuremath{\mathrm{D}}\xspace}

 \def\PK      {\ensuremath{\mathrm{K}}\xspace}

 \def\Pb      {\ensuremath{\mathrm{b}}\xspace}                 
 \def\Pc      {\ensuremath{\mathrm{c}}\xspace}

 \def\Pi      {\ensuremath{\mathrm{i}}\xspace}

 \def\Ps      {\ensuremath{\mathrm{s}}\xspace}

}
{

 \def\Ppi         {\ensuremath{\pi}\xspace}

 \mathchardef\PDelta="7101
 \mathchardef\PXi="7104
 \mathchardef\PLambda="7103
 \mathchardef\PSigma="7106
 \mathchardef\POmega="710A
 \mathchardef\PUpsilon="7107
                  
 \def\PB      {\ensuremath{B}\xspace}                 
                  
 \def\PD      {\ensuremath{D}\xspace}

 \def\PK      {\ensuremath{K}\xspace}

 \def\Pb      {\ensuremath{b}\xspace}                 
 \def\Pc      {\ensuremath{c}\xspace}

 \def\Pi      {\ensuremath{i}\xspace}

 \def\Ps      {\ensuremath{s}\xspace}

}

\def\squark    {\ensuremath{\Ps}\xspace}

\def\cquark    {\ensuremath{\Pc}\xspace}

\def\bquark    {\ensuremath{\Pb}\xspace}

\def\pion  {\ensuremath{\Ppi}\xspace}

\def\pip   {\ensuremath{\pion^+}\xspace}
\def\pim   {\ensuremath{\pion^-}\xspace}

\def\kaon  {\ensuremath{\PK}\xspace}
  \def\Kbar  {\kern 0.2em\overline{\kern -0.2em \PK}{}\xspace}

\def\Kz    {\ensuremath{\kaon^0}\xspace}
\def\Kzb   {\ensuremath{\Kbar^0}\xspace}
\def\KzKzb {\ensuremath{\Kz \kern -0.16em \Kzb}\xspace}
\def\Kp    {\ensuremath{\kaon^+}\xspace}
\def\Km    {\ensuremath{\kaon^-}\xspace}

\def\KpKm  {\ensuremath{\Kp \kern -0.16em \Km}\xspace}

\def\Kstarzb {\ensuremath{\Kbar^{*0}}\xspace}

  \def\Dbar    {\kern 0.2em\overline{\kern -0.2em \PD}{}\xspace}
\def\D       {\ensuremath{\PD}\xspace}

\def\Dz      {\ensuremath{\D^0}\xspace}
\def\Dzb     {\ensuremath{\Dbar^0}\xspace}
\def\DzDzb   {\ensuremath{\Dz {\kern -0.16em \Dzb}}\xspace}
\def\Dp      {\ensuremath{\D^+}\xspace}
\def\Dm      {\ensuremath{\D^-}\xspace}

\def\DpDm    {\ensuremath{\Dp {\kern -0.16em \Dm}}\xspace}
\def\Dstar   {\ensuremath{\D^*}\xspace}
\def\Dstarb  {\ensuremath{\Dbar^*}\xspace}

\def\Dstarp  {\ensuremath{\D^{*+}}\xspace}
\def\Dstarm  {\ensuremath{\D^{*-}}\xspace}

\def\Ds      {\ensuremath{\D^+_\squark}\xspace}
\def\Dsp     {\ensuremath{\D^+_\squark}\xspace}
\def\Dsm     {\ensuremath{\D^-_\squark}\xspace}

\def\B       {\ensuremath{\PB}\xspace}
  \def\Bbar    {\kern 0.18em\overline{\kern -0.18em \PB}{}\xspace}

\def\Bz      {\ensuremath{\B^0}\xspace}
\def\Bzb     {\ensuremath{\Bbar^0}\xspace}

\def\Bub     {\ensuremath{\B^-}\xspace}

\def\Bm      {\ensuremath{\Bub}\xspace}

\def\Bs      {\ensuremath{\B^0_\squark}\xspace}
\def\Bsb     {\ensuremath{\Bbar^0_\squark}\xspace}

  \def\Y#1S{\ensuremath{\PUpsilon{(#1S)}}\xspace}% no space before {...}!

\def\L {\ensuremath{\PLambda}\xspace}

\def\Lb      {\ensuremath{\L_\bquark}\xspace}

\def\Lc      {\ensuremath{\L_\cquark^+}\xspace}

\def\to                 {\ensuremath{\rightarrow}\xspace}

\def\CP                {\ensuremath{C\!P}\xspace}

\def\AT#1     {\ensuremath{A_{\mathrm{T}}^{#1}}\xspace}           % 2

\def\C#1      {\ensuremath{\mathcal{C}_{#1}}\xspace}                       % 9
\def\Cp#1     {\ensuremath{\mathcal{C}_{#1}^{'}}\xspace}                    % 7
\def\Ceff#1   {\ensuremath{\mathcal{C}_{#1}^{\mathrm{(eff)}}}\xspace}        % 9  
\def\Cpeff#1  {\ensuremath{\mathcal{C}_{#1}^{'\mathrm{(eff)}}}\xspace}       % 7
\def\Ope#1    {\ensuremath{\mathcal{O}_{#1}}\xspace}                       % 2
\def\Opep#1   {\ensuremath{\mathcal{O}_{#1}^{'}}\xspace}                    % 7

\newcommand{\tev}{\ensuremath{\mathrm{\,Te\kern -0.1em V}}\xspace}
\newcommand{\gev}{\ensuremath{\mathrm{\,Ge\kern -0.1em V}}\xspace}
\newcommand{\mev}{\ensuremath{\mathrm{\,Me\kern -0.1em V}}\xspace}
\newcommand{\kev}{\ensuremath{\mathrm{\,ke\kern -0.1em V}}\xspace}
\newcommand{\ev}{\ensuremath{\mathrm{\,e\kern -0.1em V}}\xspace}
\newcommand{\gevc}{\ensuremath{{\mathrm{\,Ge\kern -0.1em V\!/}c}}\xspace}
\newcommand{\mevc}{\ensuremath{{\mathrm{\,Me\kern -0.1em V\!/}c}}\xspace}
\newcommand{\gevcc}{\ensuremath{{\mathrm{\,Ge\kern -0.1em V\!/}c^2}}\xspace}
\newcommand{\gevgevcccc}{\ensuremath{{\mathrm{\,Ge\kern -0.1em V^2\!/}c^4}}\xspace}
\newcommand{\mevcc}{\ensuremath{{\mathrm{\,Me\kern -0.1em V\!/}c^2}}\xspace}

\def\mum  {\ensuremath{\,\upmu\rm m}\xspace}

\def\ps   {\ensuremath{{\rm \,ps}}\xspace}

\newcommand{\chisq}{\ensuremath{\chi^2}\xspace}

\def\gsim{{~\raise.15em\hbox{$>$}\kern-.85em
          \lower.35em\hbox{$\sim$}~}\xspace}
\def\lsim{{~\raise.15em\hbox{$<$}\kern-.85em
          \lower.35em\hbox{$\sim$}~}\xspace}

\def\pt         {\mbox{$p_{\rm T}$}\xspace}

\def\evtgen     {\mbox{\textsc{EvtGen}}\xspace}
\def\pythia     {\mbox{\textsc{Pythia}}\xspace}

\def\geant      {\mbox{\textsc{Geant4}}\xspace}

\def\photos     {\mbox{\textsc{Photos}}\xspace}

\def\tell1  {TELL1\xspace}
\def\ukl1   {UKL1\xspace}

\def\dzdzb{{~\raise.85em\hbox{{\tiny{(}\textemdash\tiny{)}}}\kern-1.05em
          \lower0.0em\hbox{$D^0$}~}\xspace}
\def\bsbsb{{~\raise.85em\hbox{{\tiny{(}\textemdash\tiny{)}}}\kern-1.05em
          \lower0.0em\hbox{$B_s^0$}~}\xspace}
\def\br{{\cal{B}}}

\def\Lb{\Lambda_b^0}

\def\eff{\epsilon}

\def\ifb{\rm fb^{-1}}

\def\btodzds{\Bm\to\Dz\Dsm}
\def\btodd{\Bzb\to\Dp\Dm}

\def\bstodsds{\Bsb\to\Dsp\Dsm}
\def\bstodd{\Bsb\to\Dp\Dm}
\def\bstodsd{\Bsb\to\Dsp\Dm}
\def\btodsd{\Bz\to\Dsp\Dm}

\def\bstodzdz{\Bsb\to\Dz\Dzb}
\def\btodzdz{\Bzb\to\Dz\Dzb}
\def\btodzdz{\Bzb\to\Dz\Dzb}
\def\btodzdz{\Bzb\to\Dz\Dzb}
\def\btodzds{\Bm\to\Dz\Dsm}
\def\Nbstodd{N_{\bstodd}}
\def\Nbstodsds{N_{\bstodsds}}
\def\Nbstodsd{N_{\bstodsd}}
\def\Nbtodsd{N_{\btodsd}}

\def\Nbtodd{N_{\btodd}}
\def\Nbtodzds{N_{\btodzds}}
\def\Nbstodzdz{N_{\bstodzdz}}
\def\Nbtodzdz{N_{\btodzdz}}
\def\rstat{{\rm stat}}
\def\rsyst{{\rm syst}}

\def\erel{\epsilon_{\rm rel}}
\def\eff{\epsilon}

\usepackage{cite} % Allows for ranges in citations
\usepackage{mciteplus}

\usepackage{longtable} % only for template; not usually to be used in PAPERs

\begin{document}

%%%%%%%%%%%%%%%%%%%%%%%%%
%%%%% Title     %%%%%%%%%
%%%%%%%%%%%%%%%%%%%%%%%%%
\renewcommand{\thefootnote}{\fnsymbol{footnote}}
\setcounter{footnote}{1}

% %%%%%%% CHOOSE TITLE PAGE--------
%\onecolumn
% \input{title-LHCb-ANA}
%\input{title-LHCb-CONF}
% $Id: title-LHCb-PAPER.tex 31008 2013-02-05 23:29:27Z sblusk $
% ===============================================================================
% Purpose: LHCb-PAPER journal paper title page template
% Author: 
% Created on: 2010-09-25
% ===============================================================================

%%%%%%%%%%%%%%%%%%%%%%%%%
%%%%%  TITLE PAGE  %%%%%%
%%%%%%%%%%%%%%%%%%%%%%%%%
\begin{titlepage}
\pagenumbering{roman}

% Header ---------------------------------------------------
\vspace*{-1.5cm}
\centerline{\large EUROPEAN ORGANIZATION FOR NUCLEAR RESEARCH (CERN)}
\vspace*{0.5cm}
\hspace*{-0.5cm}
\begin{tabular*}{\linewidth}{lc@{\extracolsep{\fill}}r}
\ifthenelse{\boolean{pdflatex}}% Logo format choice
{\vspace*{-2.7cm}\mbox{\!\!\!\includegraphics[width=.14\textwidth]{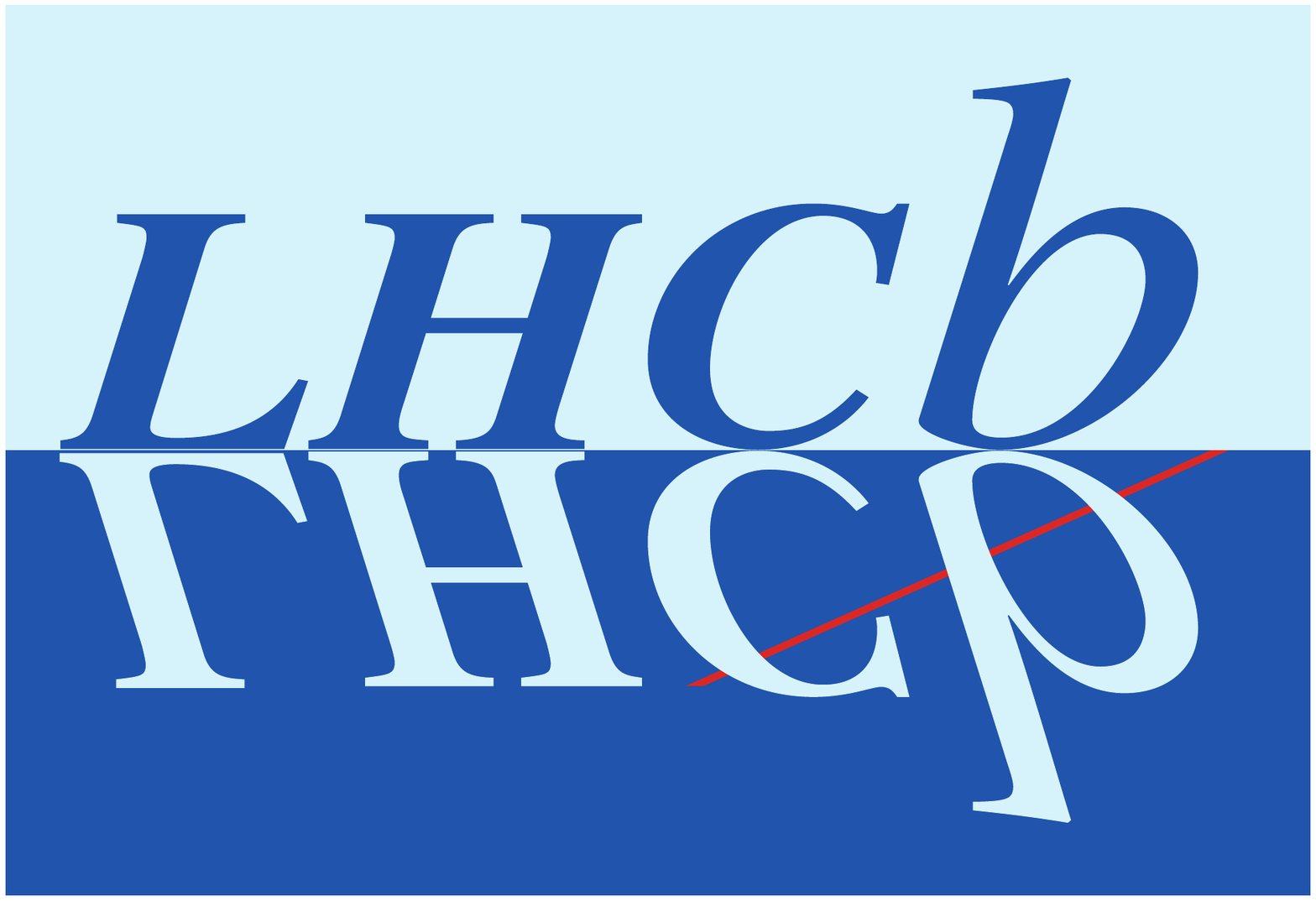}} & &}%
{\vspace*{-1.2cm}\mbox{\!\!\!\includegraphics[width=.12\textwidth]{lhcb-logo.eps}} & &}%
\\
 & & CERN-PH-EP-2013-018 \\  % ID 
 & & LHCb-PAPER-2012-050 \\  % ID 
 & & Feb. 23, 2013 \\ % Date - Can also hardwire e.g.: 23 March 2010
 & & \\
% not in paper \hline
\end{tabular*}

\vspace*{-0.1cm}

% Title --------------------------------------------------
{\bf\boldmath\huge
\begin{center}
   First observations of $\bstodd$, $\Dsp\Dm$ and $\Dz\Dzb$ decays
%and measurements of $\br(\bstodsds)$ and $\br(\btodzds)$ 
\end{center}
}
\vspace*{0.05cm}

% Authors -------------------------------------------------
\begin{center}
The LHCb collaboration\footnote{Authors are listed on the following pages.}
\end{center}

\vspace{\fill}

% Abstract -----------------------------------------------
\begin{abstract}
  \noindent
  First observations and measurements of the branching fractions of the
$\bstodd$, $\bstodsd$ and $\bstodzdz$ decays are presented using $1.0$~$\ifb$ of data collected by the LHCb experiment.
These branching fractions are normalized to those of $\btodd$, $\btodsd$ and $\btodzds$, respectively. 
An excess of events consistent with the decay $\btodzdz$ is also seen, and its branching fraction 
is measured relative to that of $\btodzds$. Improved measurements of the branching fractions
$\br(\bstodsds)$ and $\br(\btodzds)$ are reported, each relative to $\br(\btodsd)$.
The ratios of branching fractions are 
\vspace{-0.2in}
\begin{center}
\begin{align*}
{\br(\bstodd)\over \br(\btodd)} &= 1.08\pm 0.20\pm0.10, \nonumber \\
{\br(\bstodsd)\over \br(\btodsd)} &= 0.050\pm 0.008\pm0.004, \nonumber\\
{\br(\bstodzdz)\over \br(\btodzds)} &= 0.019\pm 0.003\pm0.003,  \\
{\br(\btodzdz)\over \br(\btodzds)} &= 0.0014\pm 0.0006\pm0.0002, \\ %~~( <0.0024~\rm{at}~90\%~\rm{CL} )  \nonumber \\ 
{\br(\bstodsds)\over \br(\btodsd)} &= 0.56\pm 0.03\pm0.04,  \nonumber \\
{\br(\btodzds)\over \br(\btodsd)} &= 1.22\pm 0.02\pm0.07,   \nonumber
\end{align*}
\end{center}
\noindent where the uncertainties are statistical and systematic, respectively.

\end{abstract}

\vspace*{-0.1cm}

\begin{center}
  Submitted to Physical Review D
\end{center}

\vspace{\fill}

{\footnotesize
\centerline{\copyright~CERN on behalf of the \lhcb collaboration, license \href{http://creativecommons.org/licenses/by/3.0/}{CC-BY-3.0}.}}
\vspace*{2mm}

\end{titlepage}

%%%%%%%%%%%%%%%%%%%%%%%%%%%%%%%%
%%%%%  EOD OF TITLE PAGE  %%%%%%
%%%%%%%%%%%%%%%%%%%%%%%%%%%%%%%%

%  empty page follows the title page ----
\newpage
\setcounter{page}{2}
\mbox{~}
\newpage

% Author List ----------------------------
%  You need to get a new author list!
%%%%%%%%%%%%%%%%%%%%%%%%%%%%%%%%%%%%%%%%%%
\centerline{\large\bf LHCb collaboration}
\begin{flushleft}
\small
R.~Aaij$^{40}$, 
C.~Abellan~Beteta$^{35,n}$, 
B.~Adeva$^{36}$, 
M.~Adinolfi$^{45}$, 
C.~Adrover$^{6}$, 
A.~Affolder$^{51}$, 
Z.~Ajaltouni$^{5}$, 
J.~Albrecht$^{9}$, 
F.~Alessio$^{37}$, 
M.~Alexander$^{50}$
% ...

{\footnotesize \it
% ...
$ ^{5}$Clermont Universit\'{e}, Universit\'{e} Blaise Pascal, CNRS/IN2P3, LPC, Clermont-Ferrand, France\\
$ ^{6}$CPPM, Aix-Marseille Universit\'{e}, CNRS/IN2P3, Marseille, France\\
% ...
$ ^{9}$Fakult\"{a}t Physik, Technische Universit\"{a}t Dortmund, Dortmund, Germany\\
% ...
$ ^{35}$Universitat de Barcelona, Barcelona, Spain\\
$ ^{36}$Universidad de Santiago de Compostela, Santiago de Compostela, Spain\\
$ ^{37}$European Organization for Nuclear Research (CERN), Geneva, Switzerland\\
% ...
$ ^{40}$Nikhef National Institute for Subatomic Physics, Amsterdam, The Netherlands\\
% ...
$ ^{45}$H.H. Wills Physics Laboratory, University of Bristol, Bristol, United Kingdom\\
% ...
$ ^{50}$School of Physics and Astronomy, University of Glasgow, Glasgow, United Kingdom\\
$ ^{51}$Oliver Lodge Laboratory, University of Liverpool, Liverpool, United Kingdom\\
% ...
$ ^{n}$LIFAELS, La Salle, Universitat Ramon Llull, Barcelona, Spain\\
% ...
}
\end{flushleft}
%%%%%%%%%%%%%%%%%%%%%%%%%%%%%%%%%%%%%%%%%%

\cleardoublepage

%\twocolumn
% %%%%%%%%%%%%% ---------

\renewcommand{\thefootnote}{\arabic{footnote}}
\setcounter{footnote}{0}

%%%%%%%%%%%%%%%%%%%%%%%%%%%%%%%%
%%%%%  Table of Content   %%%%%%
%%%%%%%%%%%%%%%%%%%%%%%%%%%%%%%%
%%%% Uncomment next 2 lines if desired
%\tableofcontents
%\cleardoublepage

%%%%%%%%%%%%%%%%%%%%%%%%%
%%%%% Main text %%%%%%%%%
%%%%%%%%%%%%%%%%%%%%%%%%%

\pagestyle{plain} % restore page numbers for the main text
\setcounter{page}{1}
\pagenumbering{arabic}

%% Uncomment during review phase. 
%% Comment before a final submission.
\linenumbers

% You can include short sections directly in the main tex file.
% However, for larger papers it is desirable to split the text into
% several semiautonomous files, which can be revised independently.
% This is especially useful when developing a document in
% collaboration with several people, since then different parts can be
% edited independently.  This type of file organization is shown here.
% 

\section{Introduction}
Double-charm decays of $B$ mesons can be used to probe the Cabibbo-Kobayashi-Maskawa matrix~\cite{Cabibbo:1963yz,Kobayashi:1973fv} 
elements, and provide a laboratory to study final state interactions. 
The time-dependent \CP asymmetry in the $\Bz\to\Dp\Dm$ decay provides a way to measure the
$\Bz$ mixing phase~\cite{Aubert:2008ah,Fratina:2007zk}, where information from other double-charm final states can be used to account 
for loop (penguin) contributions and other non-factorizable 
effects~\cite{Aleksan:1993qk, Sanda:1996pm, Xing:1998ca, Xing:1999yx, Pham:1999fy}.
Double-charm decays of $B$ mesons  
can also be used to measure the weak phase $\gamma$,
assuming $U$-spin symmetry~\cite{Datta:2003va, Fleischer:2007zn}. 
The purely \CP-even $\bstodsds$ decay is also of interest, as it can be used to measure the 
$\Bs$ mixing phase. Moreover, a lifetime measurement using the $\bstodsds$ decay 
provides complementary information
on $\Delta\Gamma_s$~\cite{Fleischer:2007zn, DeBruyn:2012wj, Fleischer:2011cw}
 to that obtained from direct measurements~\cite{hfag}, or
from lifetime measurements in other \CP eigenstates~\cite{Aaij:2012ns,Aaij:2012nt}.

The study of $B\to D\Dbar^{\prime}$ decays\footnote{Throughout this paper, the notation 
$D$ is used to refer to a $\Dp$, $\Dz$ or 
$\Dsp$ meson, and $B$ represents either a $\Bz$, $\Bm$ or $\Bs$ meson.} can also 
provide a better theoretical understanding of 
the processes that contribute to $B$ meson decay. Feynman diagrams contributing to the decays 
considered in this paper are shown in Fig.~\ref{fig:feyn}. 
The $\bstodzdz$, $\bstodd$ and $\btodzdz$ decays are mediated
by the $W$-exchange amplitude, 
along with penguin-annihilation contributions and rescattering~\cite{rosnerDsDs}. The only other observed $B$ meson decays
of this type are $\Bzb\to D_s^{(*)+}K^{(*)-}$ and $\Bsb\to\pip\pim$, with
branching fractions of the order of $10^{-5}$~\cite{Beringer:1900zz} and 
$10^{-6}$~\cite{Aaij:2012as}, respectively. Predictions of the $\bstodd$
branching fraction using perturbative approaches yield $3.6\times10^{-3}$~\cite{Li:2003az},
while the use of non-perturbative approaches has led to a smaller value of $1\times10^{-3}$~\cite{Eeg:2003yq}.
More recent phenomenological studies, which assume a dominant contribution
from rescattering, predict a significantly lower branching fraction of
$\br(\bstodd)=\br(\bstodzdz)=(7.8\pm4.7)\times10^{-5}$~\cite{rosnerDsDs}.

This paper reports the first observations of the $\bstodd$, $\bstodsd$ and $\bstodzdz$ decays, and
measurements of their branching fractions normalized relative to those of $\btodd$,
$\btodsd$ and $\btodzds$, respectively. An excess of events consistent with $\btodzdz$ is also seen, and its
branching fraction is reported. Improved measurements of the ratios of branching fractions 
$\br(\bstodsds)/\br(\btodsd)$ and $\br(\btodzds)/\br(\btodsd)$ are also presented. 
All results are based upon an integrated luminosity of 1.0~$\ifb$ of $pp$ collision data at
$\sqrt{s}=7$~\tev recorded by the LHCb experiment in 2011. Inclusion of charge conjugate final states is implied throughout.

\begin{figure}[t]
\begin{center}
\includegraphics[width=0.45\linewidth]{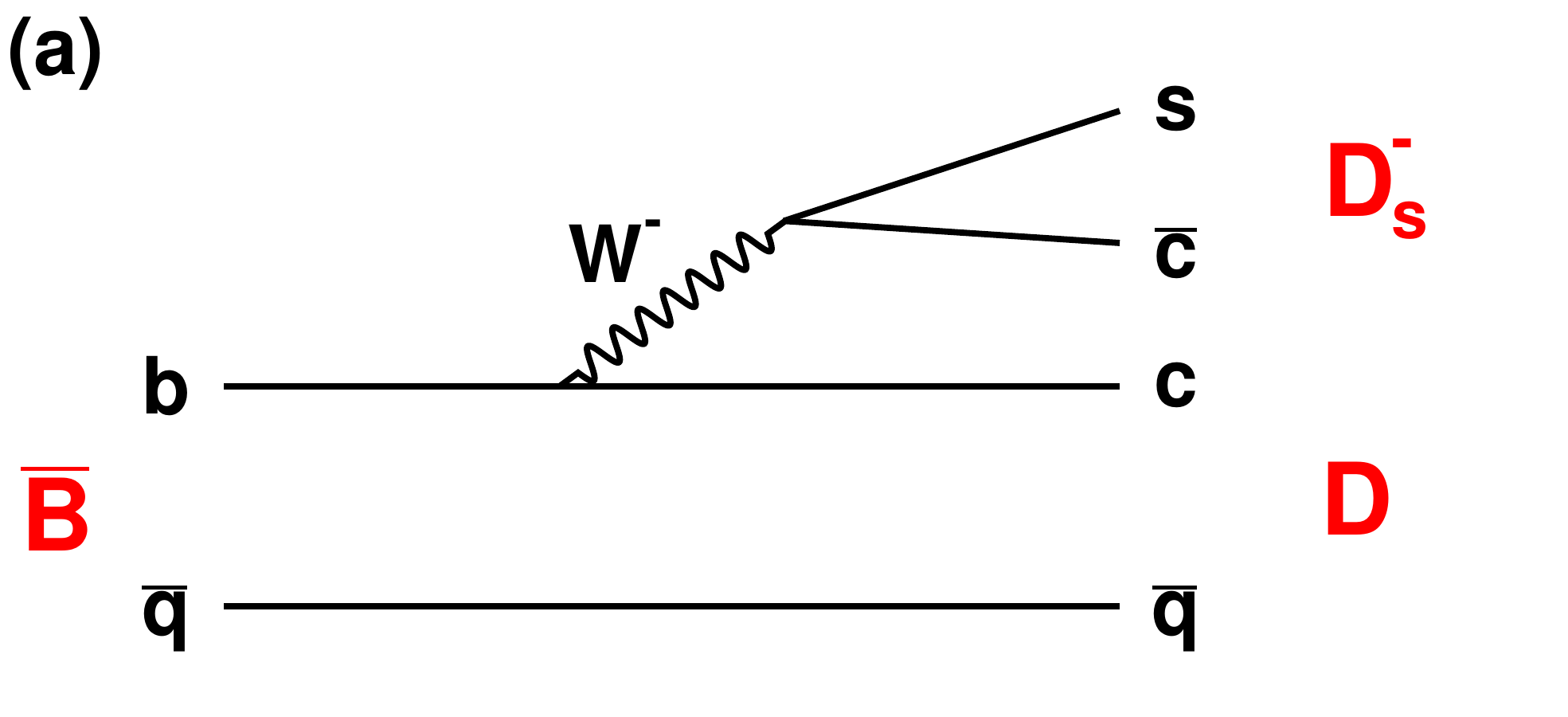}
\includegraphics[width=0.45\linewidth]{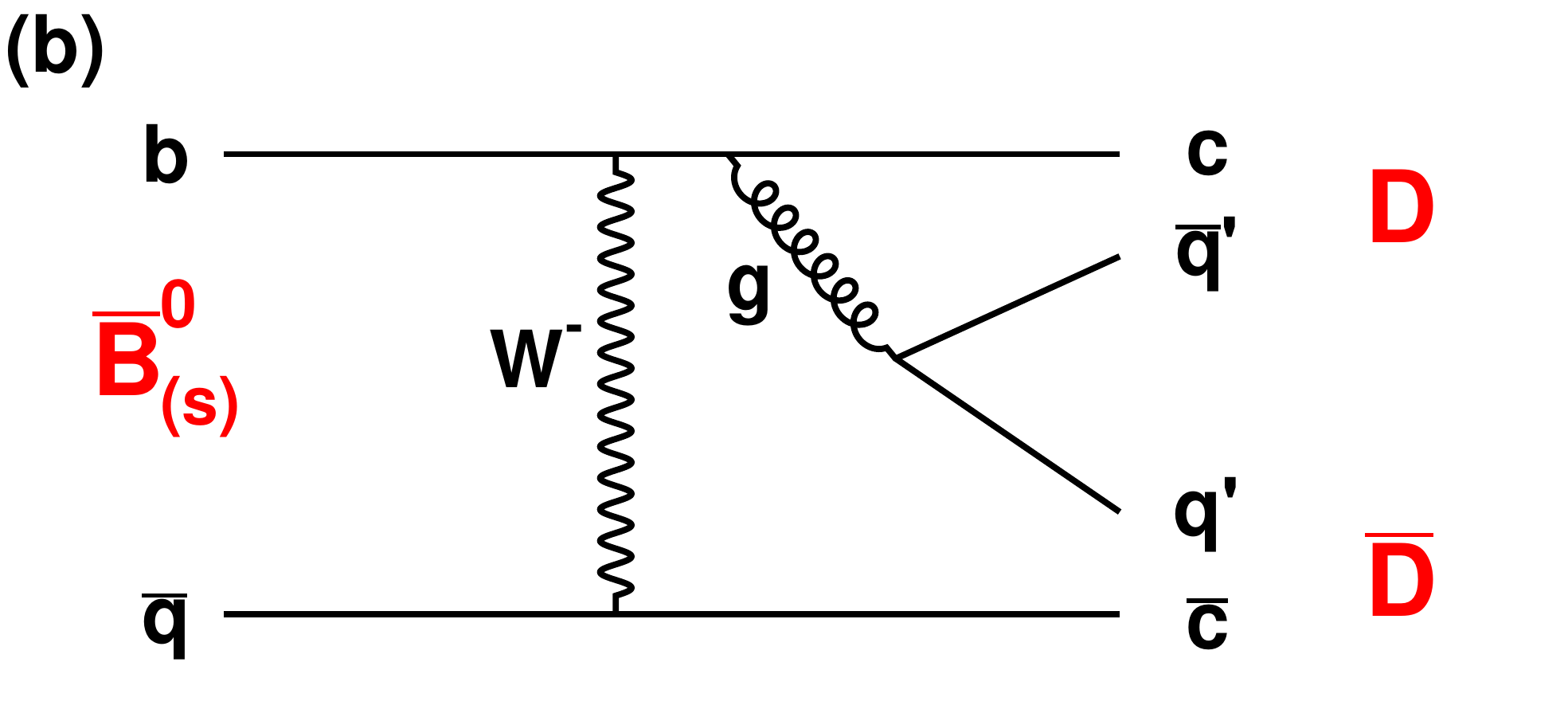}
\end{center}
\begin{center}
\includegraphics[width=0.45\linewidth]{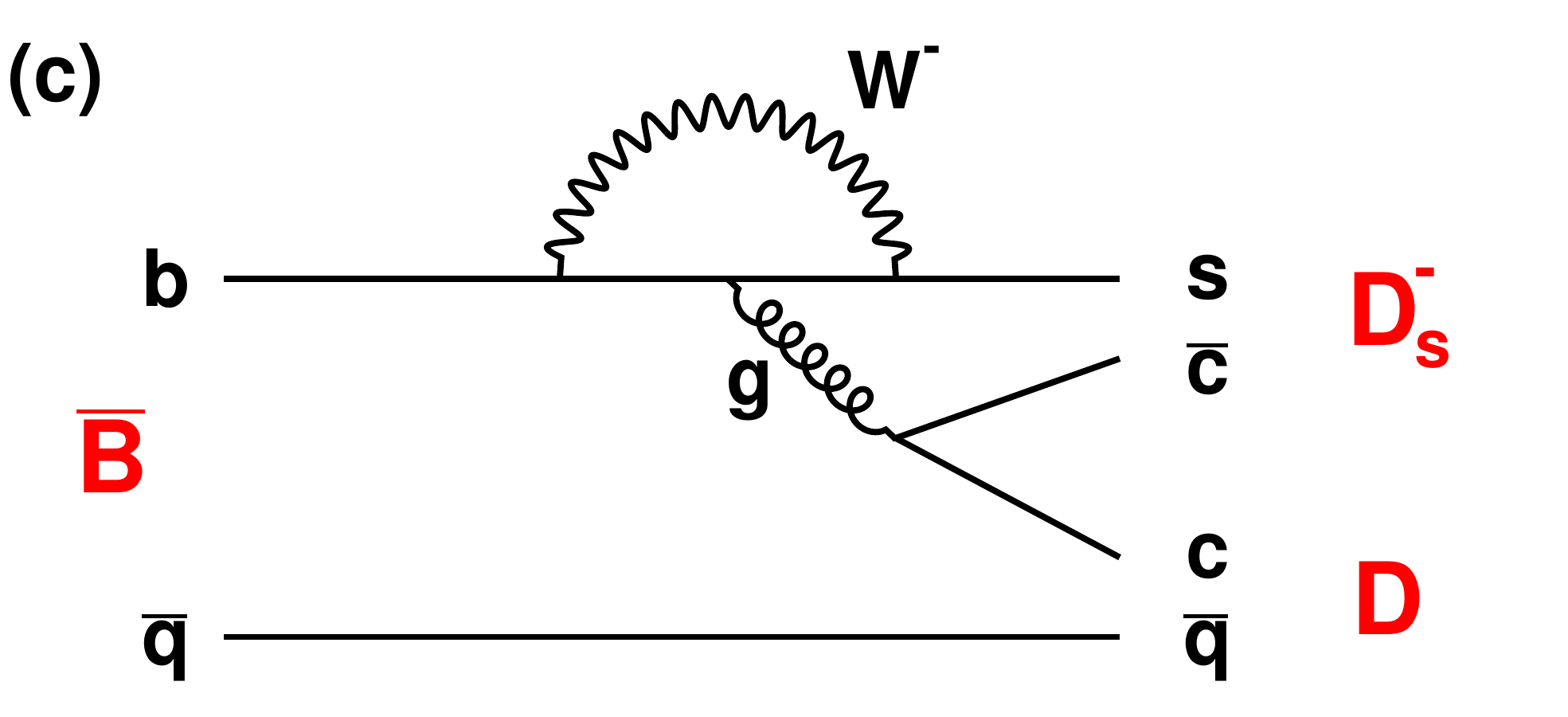}
\end{center}
\caption{\small{Feynman diagrams contributing to the double-charm final states discussed in this paper. 
They include (a) tree, (b) $W$-exchange and (c) penguin diagrams.}}
\label{fig:feyn}
\end{figure}

\section{Data sample and candidate selection}
\label{sec:selection}

The \lhcb detector~\cite{Alves:2008zz} is a single-arm forward
spectrometer covering the \mbox{pseudorapidity} range $2<\eta <5$,
designed for the study of particles containing \bquark or \cquark
quarks. The detector includes a high precision tracking system
consisting of a silicon-strip vertex detector surrounding the $pp$
interaction region, a large-area silicon-strip detector located
upstream of a dipole magnet with a bending power of about
$4{\rm\,Tm}$, and three stations of silicon-strip detectors and straw
drift tubes placed downstream. The combined tracking system has a
momentum resolution ($\Delta p/p$) that varies from 0.4\% at 5\gevc to
0.6\% at 100\gevc, and an impact parameter (IP) resolution of 20\mum for
tracks with high transverse momentum ($\pt$). The impact parameter is defined
as the distance of closest approach of a given particle to the primary
$pp$ interaction vertex (PV). Charged hadrons are identified
using two ring-imaging Cherenkov detectors~\cite{LHCbRich}. Photons, electrons
and charged particles are identified by a calorimeter system consisting of
scintillating-pad and preshower detectors, an electromagnetic
calorimeter and a hadronic calorimeter. Muons are identified by a
system composed of alternating layers of iron and multiwire
proportional chambers. 

The trigger~\cite{Aaij:2012me} consists of a hardware stage, based
on information from the calorimeter and muon systems, followed by a
software stage that performs a partial event reconstruction
(only tracks with $\pt>0.5$~\gevc are reconstructed and used).
The software trigger requires a two-, three- or four-track
secondary vertex with a large track \pt sum 
and a significant displacement from any of the reconstructed PVs.
At least one track must have $\pt >
1.7\gevc$ and IP~\chisq greater than 16 with respect to all PVs. 
The IP \chisq is defined as the
difference between the \chisq of the PV reconstructed with and
without the considered particle. A multivariate algorithm~\cite{bbdt} is used to identify
secondary vertices that originate from the decays of $b$ hadrons.

For the ratios of branching fractions between modes with identical final states,
no requirements are made on the hardware trigger decision.
When the final states differ, a trigger selection is applied to facilitate the
determination of the relative trigger efficiency. The selection requires
that either (i) at least one of the tracks from the reconstructed signal decay 
is associated with energy depositions in the calorimeters that passed the hardware 
trigger requirements, or (ii) the event triggered independently of the signal decay particles, {\it e.g.,} 
on the decay products of the other $b$ hadron in the event. 
Events that do not fall into either of these two categories ($\sim$5\%) 
are discarded.

Signal efficiencies and specific backgrounds are studied using simulated events. Proton-proton
collisions are generated using \pythia~6.4~\cite{Sjostrand:2006za} with a specific \lhcb
configuration~\cite{LHCb-PROC-2010-056}.  Decays of hadronic particles
are described by \evtgen~\cite{Lange:2001uf} in which final state
radiation is generated using \photos~\cite{Golonka:2005pn}. The
interaction of the generated particles with the detector and its
response are implemented using the \geant toolkit~\cite{Allison:2006ve, *Agostinelli:2002hh} as 
described in Ref.~\cite{LHCb-PROC-2011-006}. Efficiencies for identifying $\Kp$ and ~$\pip$ mesons
are determined using $\Dstarp$ calibration data, with kinematic quantities
reweighted to match those of the signal particles~\cite{LHCbRich}.

Signal $B$ candidates are formed by combining pairs of $D$ meson candidates
reconstructed in the following decay modes: ${\Dz\to\Km\pip}$ or ${\Km\pip\pim\pip}$, 
${\Dp\to\Km\pip\pip}$ and ${\Dsp\to\Kp\Km\pip}$. The $\Dz\to\Km\pip\pim\pip$ decay
is only used for $\Bzb_{(s)}\to\Dz\Dzb$ candidates, where a single
$\Dz\to\Km\pip\pim\pip$ decay in the final state is allowed, which 
approximately doubles the total signal efficiency. A refit of signal candidates with
$D$ mass and vertex constraints is performed to improve the $B$ mass resolution. 

Due to similar kinematics of the $\Dp\to\Km\pip\pip$, $\Dsp\to\Kp\Km\pip$ and $\Lc\to p\Km\pip$ decays,
there is cross-feed between various $b$-hadron decays that have two charm particles in the
final state. Cross-feed between $\Dp$ and $\Dsp$ occurs when the $\Km\pip h^+$ invariant mass
is within 25~\mevcc ($\sim3$ times the experimental resolution) of both the $\Dp$ and $\Dsp$ masses 
under the $h^+=\pip$ and $h^+=\Kp$ hypotheses, respectively. 
In such cases, an arbitration is performed as follows: if either $|M(\Kp\Km)-m_{\phi}|<10$~\mevcc or $h^+$ satisfies a stringent 
kaon particle identification (PID) requirement, the $D$ candidate is assigned to be a $\Dsp$ meson. Conversely, if $h^+$ passes 
a stringent pion PID requirement, the $D$ candidate is taken to be a $\Dp$ meson. Candidates that do not pass either of these
selections are rejected. A similar veto is applied to $\Dp$ and $\Dsp$ decays that are consistent with the
$\Lc\to p\Km\pip$ decay hypothesis if the proton is misidentified as a $\pip$ or $\Kp$, respectively. 
The efficiencies of these $D$ selections are determined using simulated signal decays to model the 
kinematics of the decay and $\Dstarp\to\Dz\pip$ calibration data for the PID efficiencies. Their values
are given in Table~\ref{tab:sigcuts}.

To suppress contributions from non-$D\Dbar^{\prime}$ final states,
the reconstructed $D$ decay vertex is required to be downstream of the reconstructed $B$ decay vertex, and 
the $B$ and $D$ decay vertices are required to have a vertex 
separation (VS)~\chisq larger than two. Here, the VS~\chisq is the difference in \chisq between the nominal
vertex fit and a vertex fit where the $D$ is assumed to have zero lifetime.
The efficiencies of this set of requirements are obtained from simulation and are included 
in Table~\ref{tab:sigcuts}. 

To further improve the purity of the $B\to D\Dbar^{\prime}$ samples, a boosted decision tree (BDT) discriminant
is used to distinguish signal $D$ mesons from backgrounds~\cite{Narsky:2005hn,Narsky:2005hn2}. 
The BDT uses five variables for the $D$ meson and 23 for each of its children.
The variables include kinematic quantities, track quality, and vertex and PID information.
The signal and background distributions used to train the BDT are obtained from
$\Bzb\to\Dp\pi^-$, $\Bm\to\Dz\pim$ and $\Bsb\to\Dsp\pim$ decays from data. The signal
distributions are background subtracted using weights~\cite{Pivk:2004ty} obtained from
a fit to the $B$ candidate invariant mass distribution. The background distributions are 
taken from the high $B$ mass sidebands in the same data sample.

It is found that making a requirement on the product of the two $D$ meson BDT responses provides better discrimination
than applying one to each BDT response individually.
The optimal BDT requirement in each decay is chosen by maximizing
$N_{\rm S}/\sqrt{N_{\rm S}+N_{\rm B}}$. The number of signal events, $N_{\rm S}$, is computed using
the known (or estimated, if unknown) branching fractions, selection efficiencies from simulated events, 
and the BDT efficiencies from the $\Bzb\to\Dp\pim$, $\Bm\to\Dz\pim$ and $\Bsb\to\Dsp\pim$ calibration samples, 
reweighted to account for small
differences in kinematics between the calibration and signal samples. The number, $N_{\rm B}$, 
is the expected background yield for a given BDT requirement.
The efficiencies associated with the optimal BDT cut values, determined from
an independent subset of the $\B\to\D\pim$ data, are listed in Table~\ref{tab:sigcuts}.
Correlations between the BDT values for the two \D mesons are taken into account.

For the purpose of measuring $\br(\bstodsds)/\br(\btodsd)$, only loose BDT requirements 
are imposed since the expected yields are relatively large. On the other hand, for 
$\br(\bstodsd)/\br(\btodsd)$, the expected signal yield of $\bstodsd$ decays is small; in this
case both the signal and normalization modes are required to pass the same tighter BDT requirement.
The different BDT selections applied to the $\btodsd$ decay are referred to as the ``loose selection'' and the ``tight selection.''
Since the final state is identical for the tight selection, the BDT efficiency cancels in the ratio of 
branching fractions, and is not included in Table~\ref{tab:sigcuts}.

\begin{table*}[t]
\begin{center}
\caption{\small{Individual contributions to the efficiency for selecting the various $B\to D\Dbar^{\prime}$ final states.
Shown are the efficiencies to reconstruct and trigger on the final state, and to pass the charm cross-feed veto, the
VS~\chisq and BDT selection requirements. The total selection efficiency is the product of these four values.
The relative uncertainty on the selection efficiency for each decay mode
due to the finite simulation samples sizes is 2\%. 
Entries with a dash indicate that the efficiency factor is not applicable.}}
\begin{tabular}{lcccc}
\hline\hline \\[-1.85ex]
                                & \multicolumn{4}{c}{Efficiencies (\%) } \\
                                & Rec.$\times$Trig.     & Cross-feed veto & VS~\chisq  & BDT       \\
\hline\\[-1.85ex]
$\bstodsds$                                 & 0.140 & $88.4$ & $75.4$ & ~97.5 \\
$\btodsd$ (loose selection)                 & 0.130 & $77.8$ & $82.9$ & 100.0  \\
$\Bzb_{(s)}\to\Dz\Dzb,~(\Km\pip,\Kp\pim)$         & 0.447 & $-$         & $73.7$ & ~57.8 \\
$\Bzb_{(s)}\to\Dz\Dzb,~(\Km\pip,\Kp\pim\pip\pim)$   & 0.128 & $-$         & $74.6$ & ~63.6 \\
$\btodzds$                                  & 0.238 & $92.5$  & $75.0$  & ~99.2 \\
\hline\hline
\end{tabular}
\label{tab:sigcuts}
\end{center}
\end{table*}

For $\Bzb_{(s)}\to\Dz\Dzb$ candidates, a peaking background from $B\to\Dstarp\pim\to(\Dz\pip)\pim$ decays, 
where the $\pip$ is misidentified as a $\Kp$, is observed. This contribution is removed by
requiring the mass difference, $M(\Km\pip\pip)-M(\Km\pip)>150~\mevcc$,
where the $\Kp$ in the reconstructed decay is taken to be a $\pip$. 
After the final selection around $2\%$ of events in the 
$\bstodsds$ decay mode contain multiple candidates; for all other modes the multiple candidate rate is below $1\%$.
All candidates are kept for the final analysis.

\section{Signal and background shapes}

The $B\to D\Dbar^{\prime}$ signal shapes are all similar after the $D$ mass and vertex constraints.
The signal shape is parameterized as the sum of two Crystal Ball (CB) functions~\cite{Skwarnicki:1986xj},
which account for non-Gaussian tails on both sides of the signal peak.
The asymmetric shapes account for both non-Gaussian mass resolution effects (on both sides) and energy
loss due to final state radiation. The two CB shapes are constrained to have equal area and a common mean.
Separate sets of shape parameters are determined for $\btodsd$, $\bstodsds$ and $\btodzds$ using simulated signal decays.
In the fits to data, the signal shape parameters are fixed to the simulated values, except for a
smearing factor that is added in quadrature to the widths from simulation. This number is allowed to vary
independently in each fit, but is consistent with about 4.6~\mevcc across all modes, resulting in a mass resolution 
of about 9~\mevcc.
For the more rare $\Bzb_{(s)}\to\Dz\Dzb$ and $\Bzb_{(s)}\to\Dp\Dm$ decay modes,
the $\bstodsds$ signal shape parameters are used. In determining the signal significances, the signal shape is fixed to 
that for $\bstodsds$, including an additional smearing of 4.6~\mevcc. The impact of using the $\btodsd$ or $\btodzds$ 
signal shapes on the signal significances is negligible.

Several specific backgrounds contribute to the $D\Dbar^{\prime}$ mass spectra. In particular, decays such as
$B\to D^{(*)}\Dstarb$, where the $\D^*$ mesons decay through pion or photon emission, 
produce distinct structures in all decays under consideration.
The shapes of these backgrounds are derived from simulation,
which are corrected for known resolution differences
between data and simulated events, and then fixed in fits to the data. 
The relative yield of the two peaks in the characteristic structure from the decay 
$\Dstar\to\Dz\pi$ is allowed to vary freely, to enable better modeling of the background  
in the low mass region. Since this mass region is significantly below the signal peaks, the impact on the
signal yield determinations is negligible.

A source of peaking background that contributes to $B\to D\Dsp$ modes are
the ${B\to\D\Kstarzb\Kp\to D\Km\pip\Kp}$ decays, where the $\Kstarzb\Kp$ is not produced in a $\Dsp$ decay. 
Although the branching fractions for these decays~\cite{Drutskoy:2002ib} are about twice as large as that of the
$B\to\D\Ds\to\D\Kp\Km\pip$ decay channel, the 25~\mevcc mass window around the known $\Dsp$ mass and the VS~\chisq$>2$ requirement reduce
this contribution to about 1\% of the signal yield.
This expectation is corroborated by studying the $\Dsp$ candidate mass sidebands.
The shape of this background is obtained from simulation, and is described by a single Gaussian function
which has a width about 2.5 times larger than that of the signal decay and 
peaks at the nominal $B$ meson mass.

After the charm cross-feed vetoes (see Sect.~\ref{sec:selection}), the cross-feed rate from 
$\btodsd$ decays into the $\bstodsds$ sample is $(0.7\pm0.2)\%$. The shape of this misidentification background 
is obtained from simulation.
A similar cross-feed background contribution from $\Lb\to\Lc\Dsm$ decays is also expected due to events passing
the $\Lc$ veto. Taking into account the observed yields of these decays in data, we fix the $\btodsd$ and $\Lb\to\Lc\Dsm$ 
cross-feed yields to 35 and 15 events, respectively. Investigation of the $D$ mass sidebands 
reveals no additional contributions from non-$D\Dbar^{\prime}$ backgrounds.

The combinatorial background shape is described by an exponential function whose slope is determined from
wrong-sign candidates. Wrong-sign candidates include the $D_{s}^{+}D_{s}^{+}$, $\Dz\Dz$, or $\Dzb(\Kp\pim)\Dsm$ final states, in which no signal
excesses should be present (neglecting the small contribution from the doubly Cabibbo suppressed $\Bm\to\Dz(\Kp\pim)\Dsm$ decay).
For the $\Bzb_{(s)}\to\Dp\Dm$ decay, the exponential shape parameter is allowed to vary in the fit 
due to an insufficient number of wrong-sign $\Dp\Dp$ candidates.

\section{Fit results}
Figure~\ref{fig:bs2dsds} shows the invariant mass spectra for $\bstodsds$ and
$\btodsd$ candidates. The results of unbinned extended maximum likelihood fits to the distributions 
are overlaid with the signal and background components indicated in the legends. 
Signal yields of $451\pm23$ $\bstodsds$ and $5157\pm64$ $\btodsd$ decays are observed.

\begin{figure}[tb]
\centering
\includegraphics[width=0.49\textwidth]{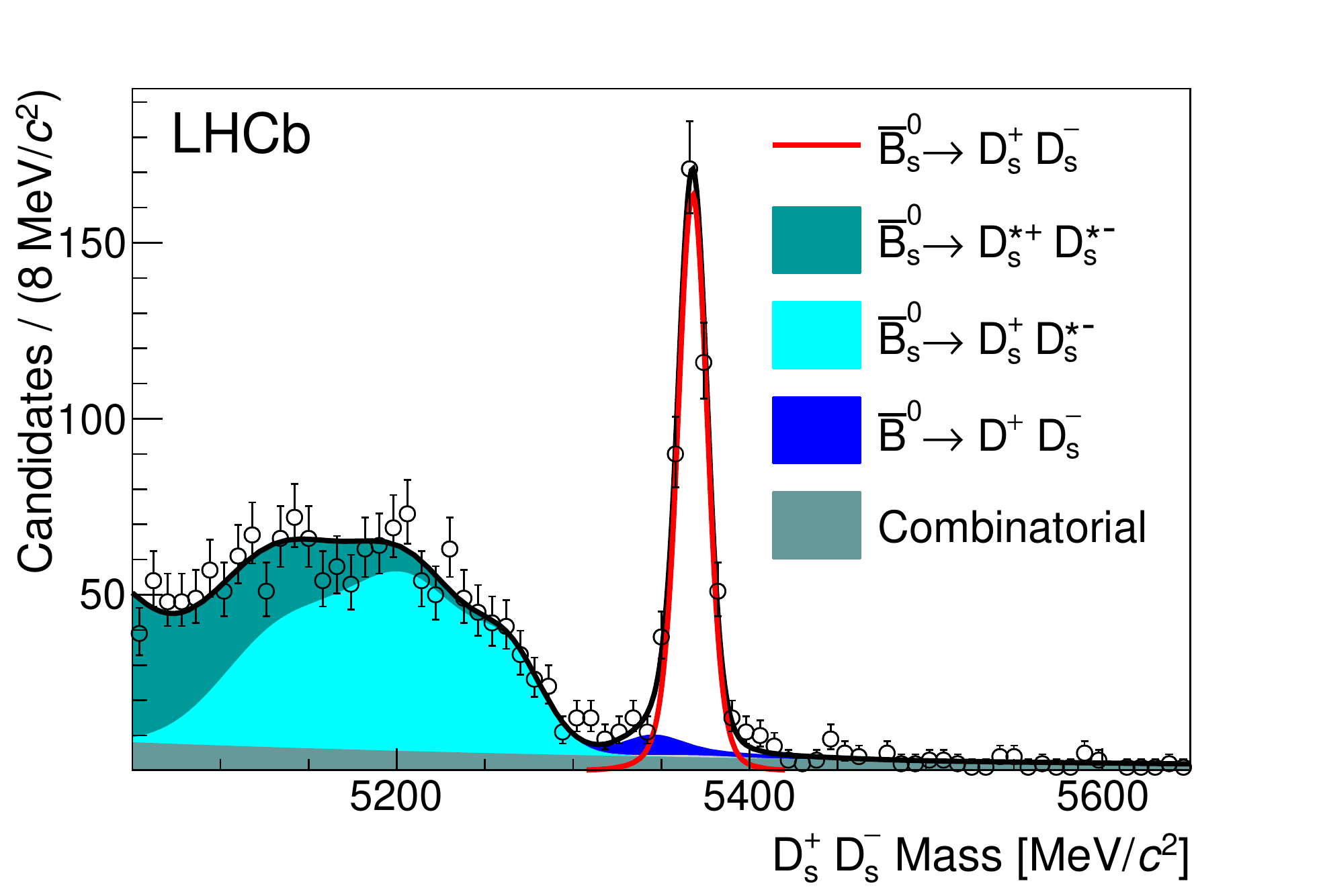}
\includegraphics[width=0.49\textwidth]{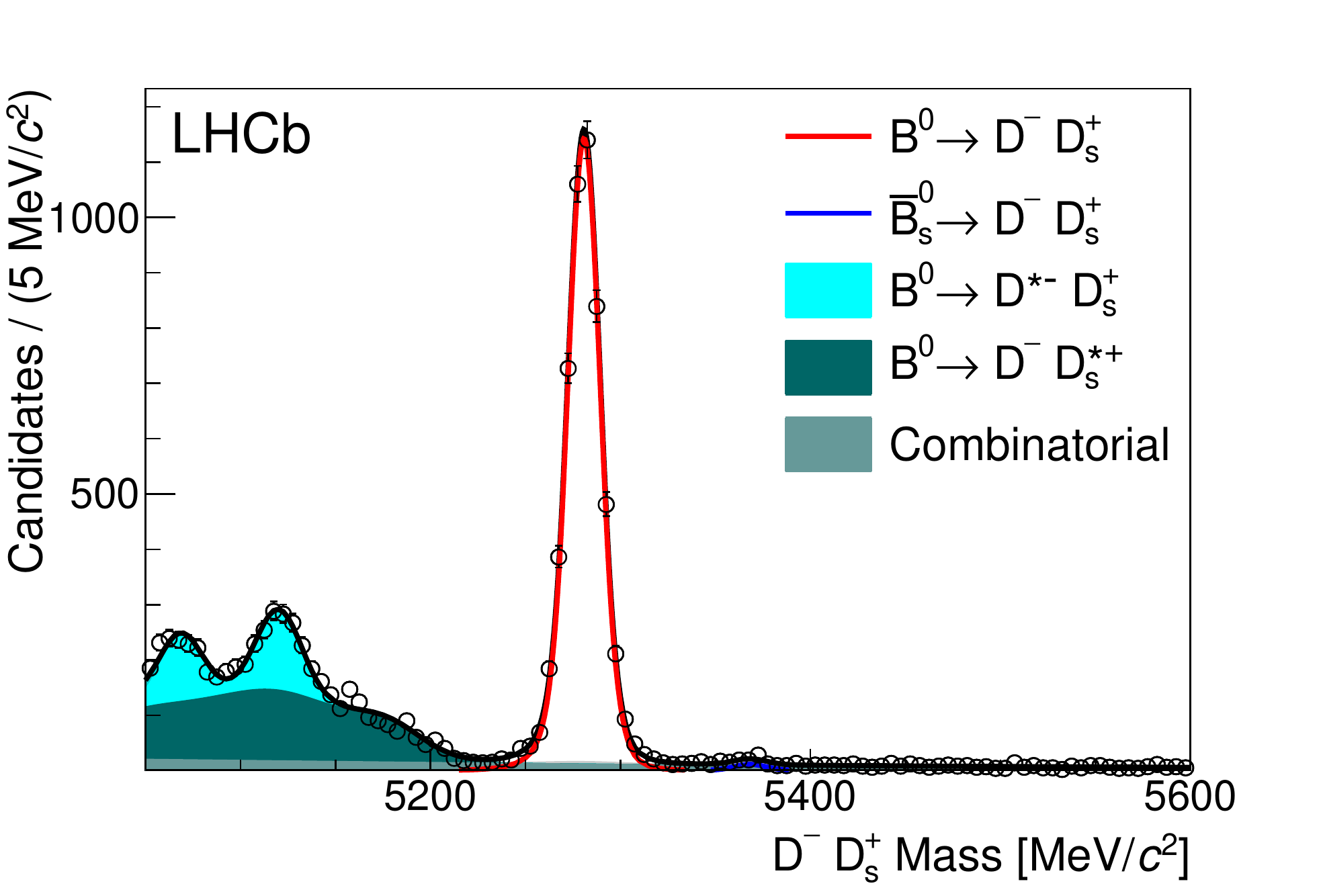}
\caption{\small{Invariant mass distributions for (left) $\bstodsds$ and  (right) $\btodsd$ candidates in the data
with the loose BDT selection applied to the latter.
The signal and background components are indicated in the legend. The $\Lb\to\Lc\Dsm$, $\Bsb\to\Dsp\Km\Kp\pim$
and $\Bz\to\Dm\Kp\Km\pip$ background components are too small to be seen, and are excluded from the legends.}}
\label{fig:bs2dsds}
\end{figure}

Figure~\ref{fig:bs2dsd} shows the invariant mass spectrum for 
$\btodsd$ and $\bstodsd$ candidates, where the tight BDT selection requirements have been applied as discussed previously.
We observe $36\pm6$ $\bstodsd$ signal decays, with $2832\pm53$ events in the $\btodsd$ normalization mode.
The statistical significance of the $\bstodsd$ signal corresponds to $10\sigma$ by computing
$\sqrt{-2\,{\rm ln}({\mathcal{L}_{0}}/{\mathcal{L}_{\rm max}}})$,
where ${\mathcal{L}_{\rm max}}$ and ${\mathcal{L}_{0}}$ are the fit likelihoods with 
the signal yields allowed to vary and fixed to zero, respectively. Variations in the signal and background model
have only a marginal impact on the signal significance.
The $\Bsb\to\Dm\Dsp$ decay is thus observed for the first time. 

\begin{figure}[tb]
\centering
\includegraphics[width=0.49\textwidth]{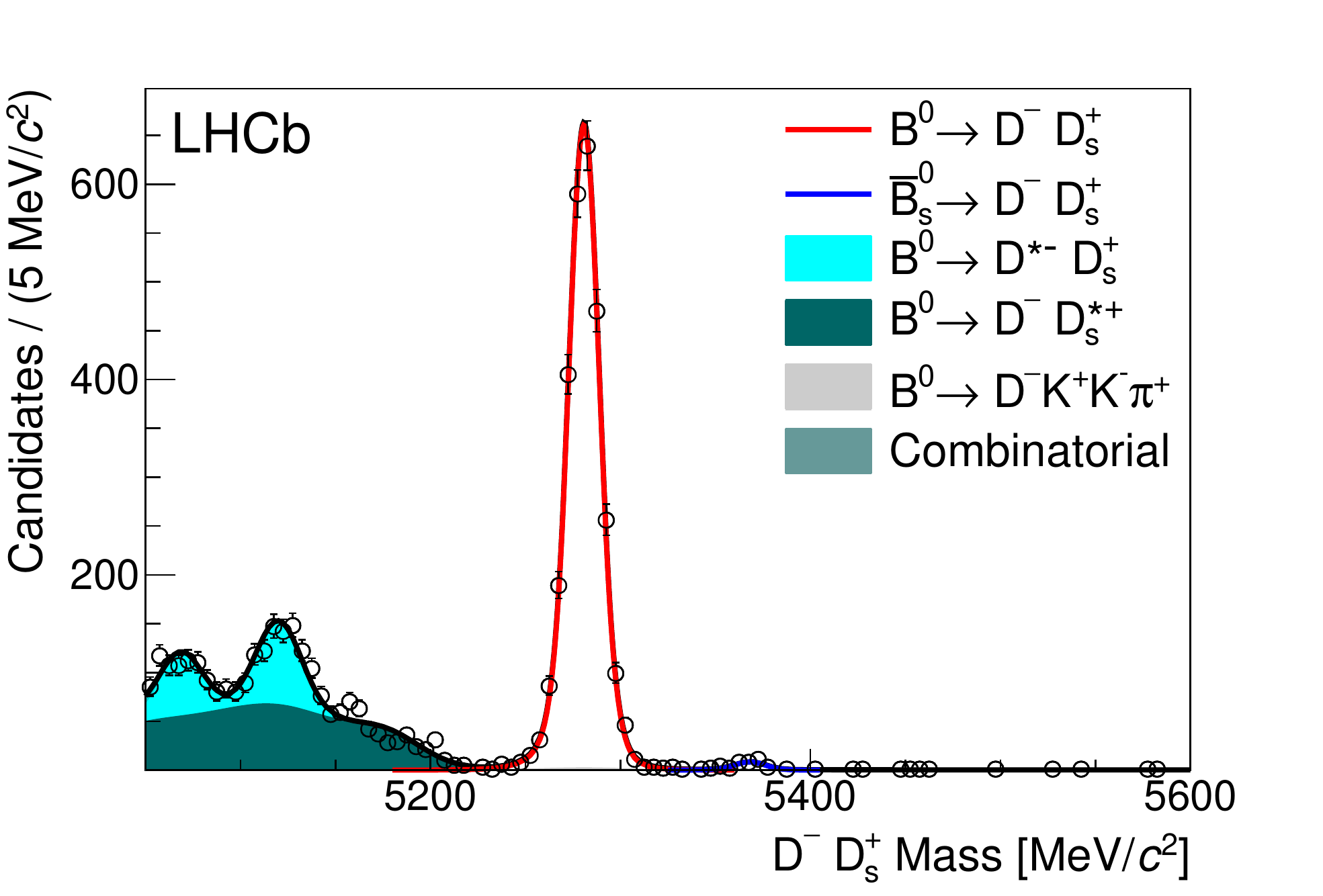}
\includegraphics[width=0.49\textwidth]{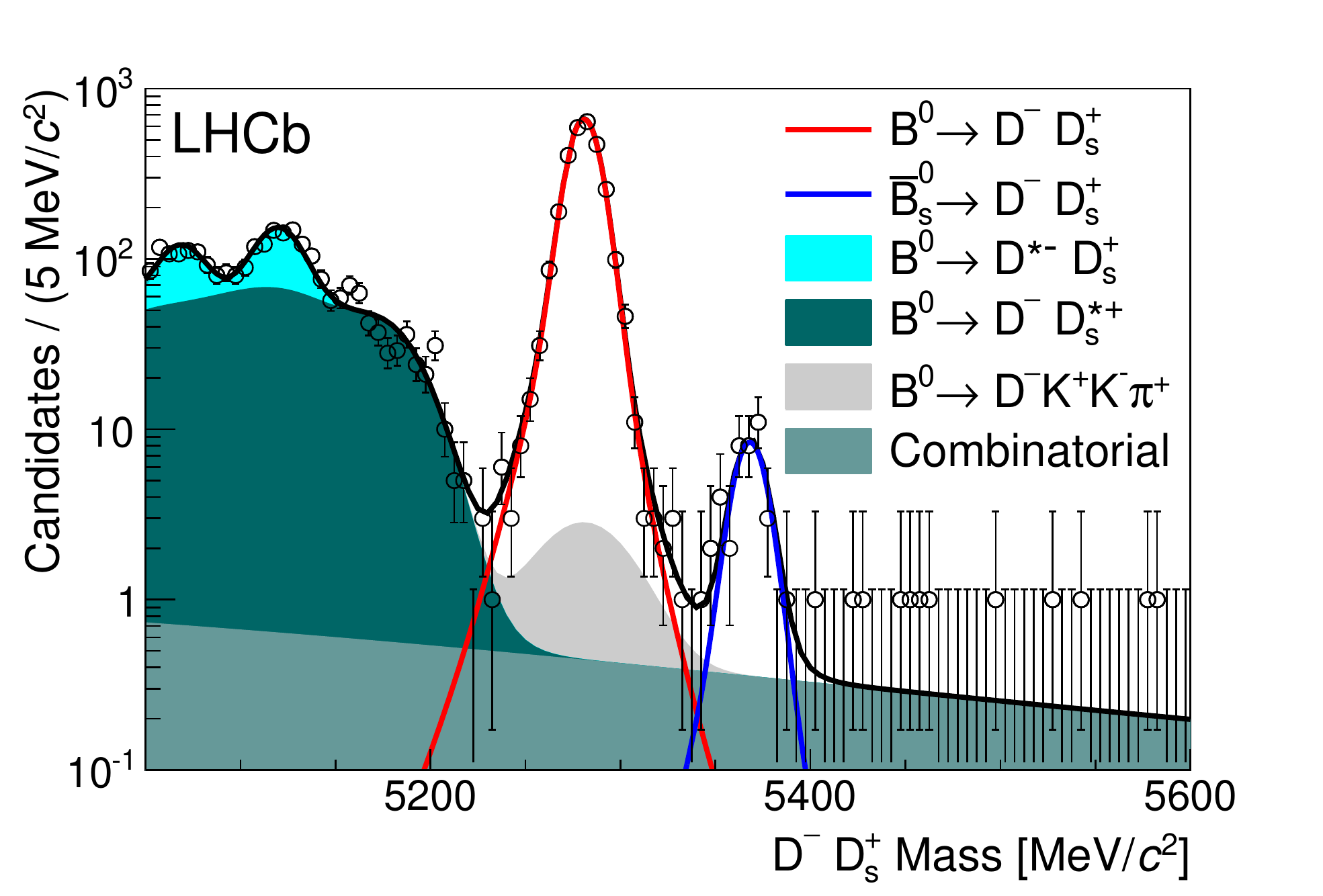}
\caption{\small{Invariant mass distribution for $\btodsd$ and $\bstodsd$
candidates in the data, with the tight BDT selection applied.
The distribution is plotted on a (left) linear and (right) logarithmic scale to highlight the suppressed $\bstodsd$ signal.
Signal and background components are indicated in the legend.}}
\label{fig:bs2dsd}
\end{figure}

The invariant mass spectrum for $\Bzb_{(s)}\to\Dp\Dm$ candidates is shown in Fig.~\ref{fig:b2dd} (left).
Peaks are seen at both the $\Bz$ and $\Bs$ meson masses, with yields of $165\pm13$ and $43\pm7$ signal 
events, respectively. In the lower mass region,
two prominent peaks from $\Bzb\to\Dstarp\Dm$ and $\Bzb\to\Dp\Dstarm$ decays are also evident. 
The significance of the $\bstodd$ signal yield
is computed as described above, and corresponds to $11\sigma$, establishing the first observation of this 
decay mode. 

Figure~\ref{fig:b2dd}\,(right) shows the $\Dz\Dzb$ invariant mass distribution and the results of the fit. 
Both ($\Km\pip$,~$\Kp\pim$) and ($\Km\pip$,~$\Kp\pim\pip\pim$) combinations are included. A $\bstodzdz$
signal is seen with a significance of 11$\sigma$, which establishes the first observation of this decay mode.
The data also show an excess of events at the $\Bz$ mass.
The significance of that excess corresponds to 2.4$\sigma$, including both the statistical and systematic uncertainty.
The fitted yields in the $\bstodzdz$ and $\btodzdz$ decay modes are $45\pm8$ and 
$13\pm6$ events, respectively. If both the $\bstodzdz$ and $\btodzdz$ decays proceed through 
$W$-exchange diagrams, one would expect the signal yield in $\btodzdz$ to be 
$\sim(f_d/f_s)\times |V_{cd}/V_{cs}|^2\simeq0.2$ of the yield in $\bstodzdz$,
where we have used $|V_{cd}/V_{cs}|^2=0.054$~\cite{Beringer:1900zz} and $f_s/f_d = 0.256\pm0.020$~\cite{LHCb-PAPER-2012-037}.
The fitted yields are consistent with this expectation.
The decay $\btodzds$ is used as the normalization channel for both the $\bstodzdz$ and $\btodzdz$ branching fraction 
measurements, where only the $\Dz\to\Km\pip$ decay mode is used. The fitted invariant mass distribution for 
$\btodzds$ candidates is shown in Fig.~\ref{fig:b2dsdz_wfdchi2cut}. The fitted signal yield is $5152\pm73$ events.

\begin{figure}[tb]
\centering
\includegraphics[width=0.49\textwidth]{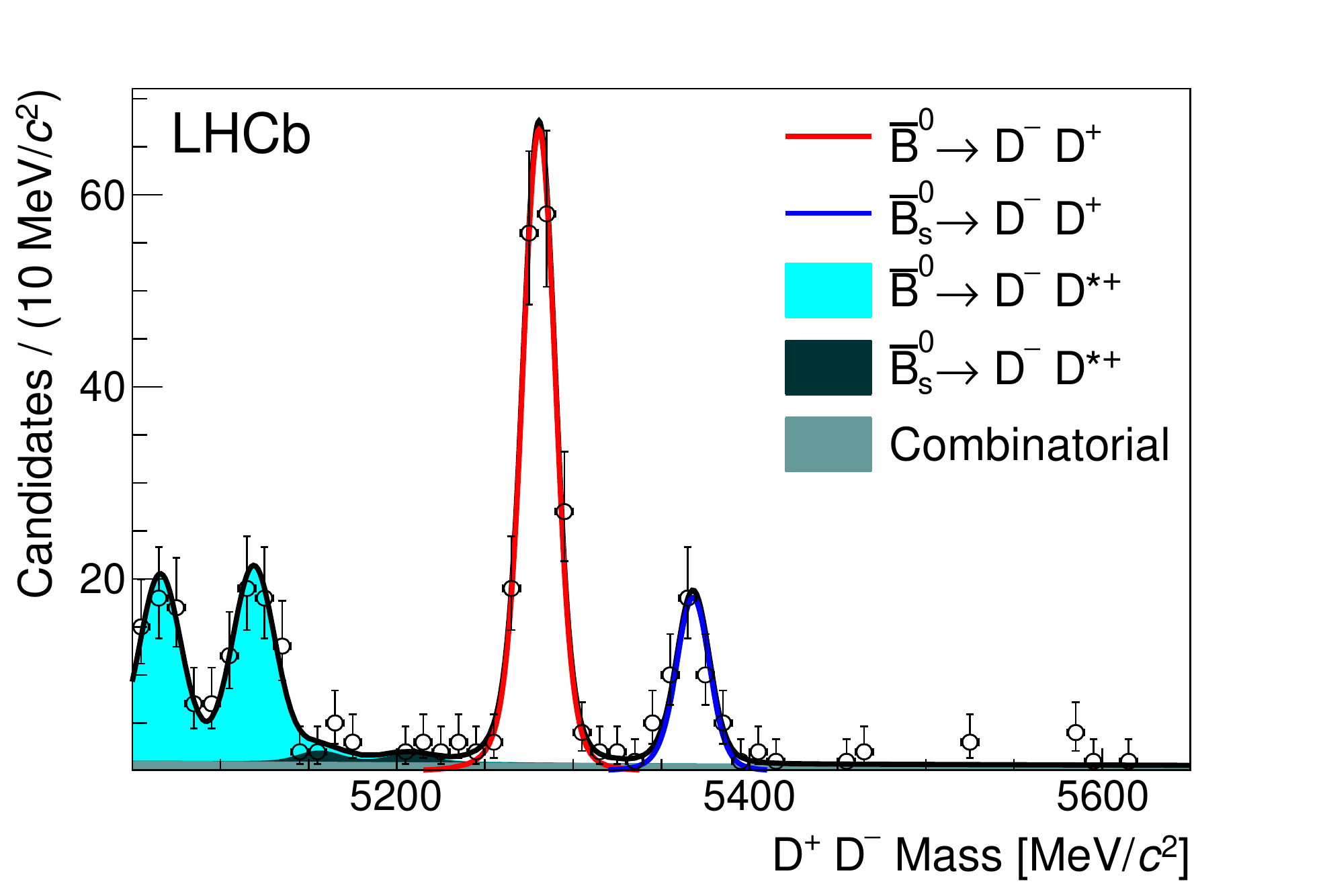}
\includegraphics[width=0.49\textwidth]{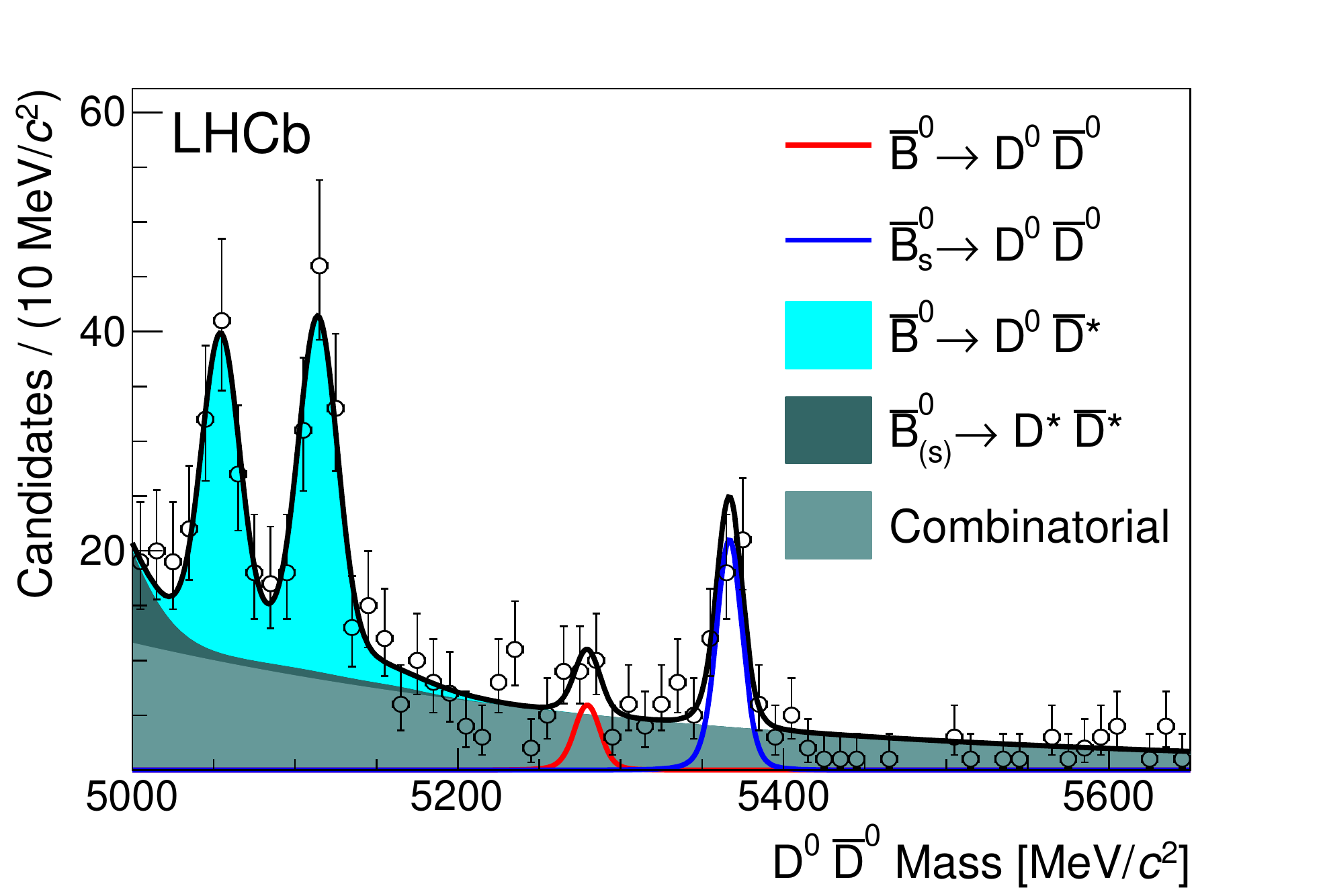}
\caption{\small{Invariant mass distributions for (left) $\Bzb_{(s)}\to\Dp\Dm$ and (right) $\Bzb_{(s)}\to\Dz\Dzb$ candidates in the data.
Signal and background components are indicated in the legend.}}
\label{fig:b2dd}
\end{figure}

\begin{figure}[tb]
\centering
\includegraphics[width=0.49\textwidth]{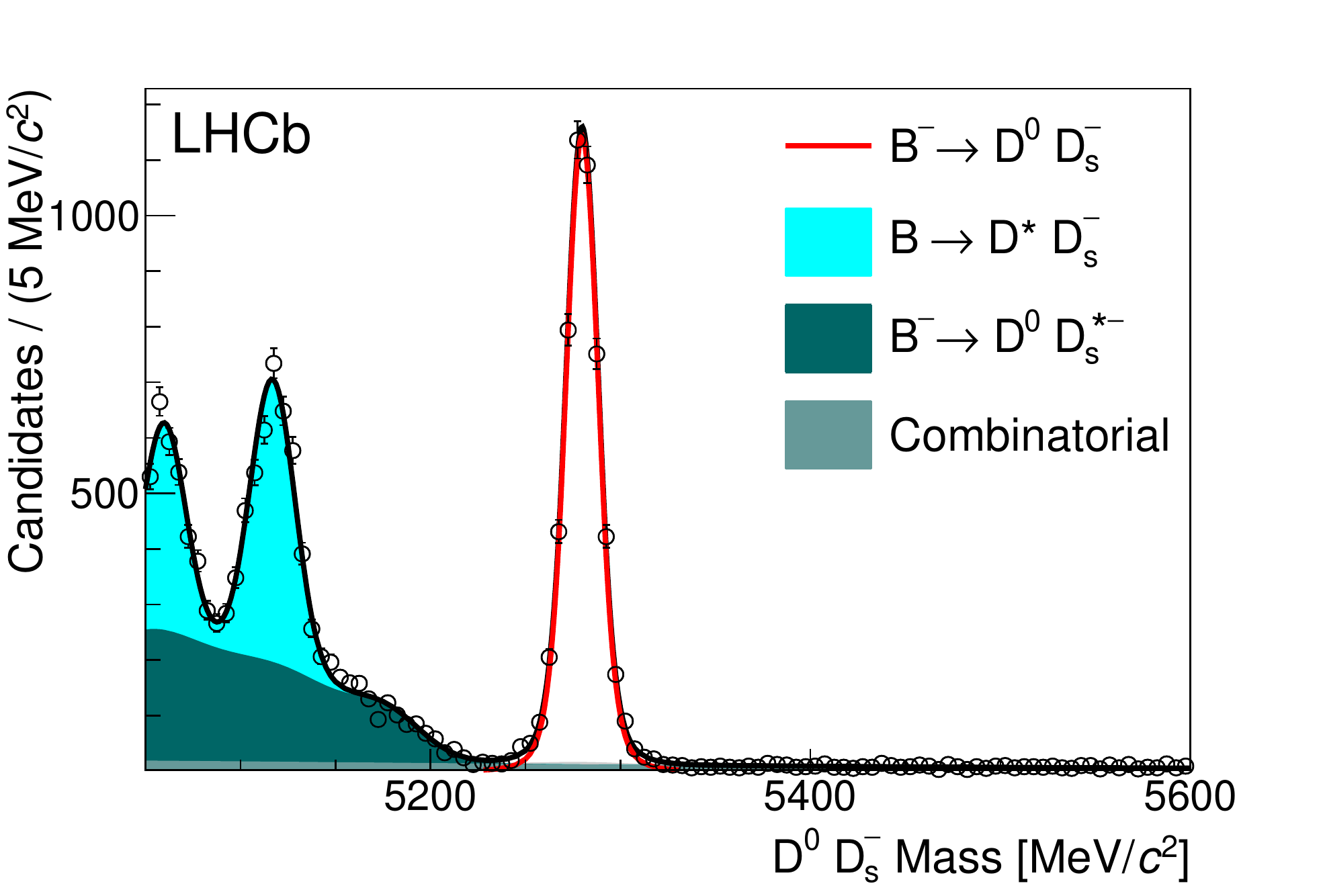}
\caption{\small{Invariant mass distribution for $\Bm\to\Dz\Dsm$ candidates in the data.
Signal and background components are indicated in the legend. The $\Bm\to\Dz\Km\Kp\pim$ background components  
are too small to be seen, and are excluded from the legend.}}
\label{fig:b2dsdz_wfdchi2cut}
\end{figure}

The measured yields, $N_{B\to D\Dbar^{\prime}}$, relevant for the branching fraction measurements are summarized in Table~\ref{tab:yields}.
The branching fractions are related to the measured yields by
\begin{align}
\label{eq:compbf2}
{\br(\bstodsds)\over \br(\btodsd)} &=  {f_d\over f_s}\cdot \erel^{\Bz/\Bs}\cdot \kappa\cdot {\br(\Dp\to\Km\pip\pip)\over \br(\Dsp\to\Kp\Km\pip)}\cdot {\Nbstodsds\over\Nbtodsd}, \\
{\br(\bstodsd)\over \br(\btodsd)} &= {f_d\over f_s}\cdot\erel\cdot {\Nbstodsd\over\Nbtodsd},   \\ 
{\br(\bstodd)\over \br(\btodd)} &= {f_d\over f_s}\cdot\erel\cdot \kappa\cdot {\Nbstodd\over\Nbtodd}, \\
\label{eq:compbf3}
{\br(\bstodzdz)\over \br(\btodzds)} &=  {f_d\over f_s}\cdot \erel^{\prime}\cdot \kappa\cdot {\Nbstodzdz\over\Nbtodzds}, \\
{\br(\btodzdz)\over \br(\btodzds)} &=  \erel^{\prime}\cdot {\Nbtodzdz\over\Nbtodzds}, 
\end{align}
%\noindent and
\begin{align}
\label{eq:compbf}
{\br(\btodzds)\over \br(\btodsd)} &=  \erel^{\Bz/\Bm}\cdot {\br(\Dp\to\Km\pip\pip)\over\br(\Dz\to\Km\pip)}\cdot {\Nbtodzds\over\Nbtodsd}~.
\end{align}
\noindent 
Here, it is assumed that $\Bm$ and $\Bzb$ mesons are produced in equal numbers.
The relative efficiencies, $\erel$, are given in Table~\ref{tab:yields}. They account
for geometric acceptance, detection and trigger efficiencies,  
and the additional VS~\chisq, BDT, and charm cross-feed veto requirements. The first four of these relative efficiencies 
are obtained from simulation, and the last two are data-driven.
The indicated uncertainties on the relative efficiencies are due only to the finite sizes of the simulated signal decays. 
The average selection efficiency for $\btodzds$ relative to $\Bzb_{(s)}\to\Dz\Dzb$ is 
\begin{align}
  \erel^{\prime} = {\eff_{\btodzds}\br(\Dsp\to\Kp\Km\pip)\br(\Dz\to\Km\pip) \over \eff_{K\pi,K\pi}[\br(\Dz\to\Km\pip)]^2+2\eff_{K\pi\pi\pi,K\pi}\br(\Dz\to\Km\pip)\br(\Dz\to\Km\pip\pim\pip)},
\label{eq:dzdz}
\end{align}
\noindent where the quantities $\eff_{\btodzds}=(0.166\pm0.003)\%$, $\eff_{K\pi,K\pi}=(0.190\pm0.003)\%$ and 
$\eff_{K\pi\pi\pi,K\pi}=(0.061\pm0.002)\%$ are the selection efficiencies
for the ${\btodzds}$, ${\Bsb\to(\Dz\to\Km\pip,\Dzb\to\Kp\pim)}$ and ${\Bsb\to(\Dz\to\Km\pip, \Dzb\to\Kp\pim\pip\pim)}$ decays, 
respectively. The $D$ branching fractions, ${\br(\Dz\to\Km\pip)=(3.88\pm0.05)\%}$, ${\br(\Dz\to\Km\pip\pim\pip)=(8.07\pm0.20)\%}$,
${\br(\Dsp\to\Kp\Km\pip)=(5.49\pm0.27)\%}$, and ${\br(\Dp\to\Km\pip\pip)=(9.13\pm0.19)\%}$ are taken from Ref.~\cite{Beringer:1900zz}. 

\begin{table*}[tb]
\begin{center}
\caption{\small{Summary of the observed signal and normalization mode yields and their relative efficiencies, as 
used in the measurements of the ratios of branching fractions.
The quoted uncertainties are statistical only.}}
\begin{tabular}{lccc}
\hline\hline
Measurement                          &  Signal   &      Norm.  & Rel. eff.      \\
                                     &  yield       &   yield & $\erel^{(\prime)}$        \\
\hline\\[-1.5ex]
${\br(\bstodsds)\over \br(\btodsd)}$ &  $451\pm23$  & $~5157\pm64~$ & $~0.928\pm0.027~$  \\
 & & & \\[-1.85ex]
${\br(\bstodsd)\over \br(\btodsd)}$  & $36\pm6$     & $~2832\pm53~$  & 1.0 \\
 & & & \\[-1.85ex]
${\br(\bstodd)\over \br(\btodd)}$    & $43\pm7$     & $~~~165\pm13~$ & 1.0 \\
 & & & \\[-1.85ex]
${\br(\bstodzdz)\over \br(\btodzds)}$ &   $45\pm8$   & $~5152\pm73~$ & $~0.523\pm0.016~$ \\
 & & & \\[-1.85ex]
${\br(\btodzdz)\over \br(\btodzds)}$ &   $13\pm6$   & $~5152\pm73~$  & $~0.523\pm0.016~$ \\
 & & & \\[-1.85ex]
${\br(\btodzds)\over \br(\btodsd)}$ &   $5152\pm73~~$   & $~5157\pm64~$ & $~0.508\pm0.011~$ \\
 & & & \\[-1.85ex]
\hline\hline
\end{tabular}
\label{tab:yields}
\end{center}
\end{table*}

The factor $\kappa$ is a correction that accounts for the lower selection efficiency associated with
the shorter-lifetime \CP-even eigenstates of the $\Bs$ system compared to flavor-specific final states~\cite{hfag}.
The impact on the $\Bs$ acceptance is estimated by convolving 
an exponential distribution that has a 10\% smaller lifetime than that in flavor-specific decays
with the simulated lifetime acceptance.
The resulting correction is $\kappa=1.058\pm0.029$. In the $\Bz$ sector, $\Delta\Gamma_d/\Gamma_d$ is 
below 1\%~\cite{Lenz:2011ti}, and the lifetime acceptance is well described by the simulation.

The measured ratios of branching fractions are computed to be
\begin{align*}
{\br(\bstodd)\over \br(\btodd)} &= 1.08\pm 0.20\,(\rstat)\pm0.10\,(\rsyst), \nonumber \\
{\br(\bstodsd)\over \br(\btodsd)} &= 0.050\pm 0.008\,(\rstat)\pm0.004\,(\rsyst), \nonumber\\
{\br(\bstodzdz)\over \br(\btodzds)} &= 0.019\pm 0.003\,(\rstat)\pm0.003\,(\rsyst),  \\
{\br(\btodzdz)\over \br(\btodzds)} &= 0.0014\pm 0.0006\,(\rstat)\pm0.0002\,(\rsyst)  \nonumber \\ 
                                   & [~< 0.0024~\rm{at}~90\%~\rm{CL}~] , \nonumber \\
{\br(\bstodsds)\over \br(\btodsd)} &= 0.56\pm 0.03\,(\rstat)\pm0.04\,(\rsyst),  \nonumber \\
{\br(\btodzds)\over \br(\btodsd)} &= 1.22\pm 0.02\,(\rstat)\pm0.07\,(\rsyst).   \nonumber
\end{align*}
For ${\br(\bstodzdz)/\br(\btodzds)}$, the results obtained using the 
$\Dz(\Km\pip)\Dzb(\Kp\pim\pip\pim)$ and $\Dz(\Km\pip)\Dzb(\Kp\pim)$ final states differ
by less than one standard deviation. 
For the $\btodzdz$ decay, we provide both the central value and the 90\% confidence level  (CL) upper limit.
The upper limit is obtained by convolving the fitted likelihood with a Gaussian function whose width
is the total systematic error, and integrating over the physical region.

\section{Systematic uncertainties}

A number of systematic uncertainties contribute to the measurements of the ratios of branching fractions.
The sources and their values are summarized in Table~\ref{tab:syst}. 
The dominant source of uncertainty on the branching fraction ratios comes from the $b$ fragmentation fraction ratio, 
$f_d/f_s$, which has a total uncertainty of 7.8\%~\cite{LHCb-PAPER-2012-037}, of which
5.3\% is from the ratio of branching fractions $\br(\Dsp\to\Kp\Km\pip)/\br(\Dp\to\Km\pip\pip)$. For clarity, we have
removed that portion of the uncertainty from $f_d/f_s$, and included its contribution in the row labeled $\br(D)$
in Table~\ref{tab:syst}. For $\br(\bstodsds)/\br(\btodsd)$, the above $\Dsp/\Dp$ branching fraction ratio from $f_d/f_s$
cancels with the corresponding inverted ratio in Eq.~\ref{eq:compbf2}. On the other hand, in the ratio
$\br(\Bzb_{(s)}\to\Dz\Dzb)/\br(\btodzds)$, the $\Ds\to\Kp\Km\pip$ branching fraction
enters as the square, after considering the $D$ branching fractions used in computing $f_d/f_s$ 
(see Eq.~\ref{eq:compbf3}). 
As a result, the uncertainty from $\br(\Dsp\to\Kp\Km\pip)$  contributes 9.8\% to the total uncertainty
on $\br(\Bzb_{(s)}\to\Dz\Dzb)/\br(\btodzds)$; smaller contributions from the limited
knowledge of $\br(\Dz\to\Km\pip)$ [1.3\%], $\br(\Dz\to\Km\pip\pim\pip)$ [2.5\%] and $\br(\Dp\to\Km\pip\pip)$ [2.1\%]
are also included in the $\br(D)$ uncertainties.

Another significant uncertainty results from the precision on $b$-hadron lifetimes and decays of
$\Bz$ and $\Bs$ to \CP eigenstates. Using the measured value of the width difference,
$\Delta\Gamma_s=0.116\pm0.018\pm0.006~\ps^{-1}$~\cite{LHCb-CONF-2012-002} 
we conservatively assume the \CP-even lifetime to be in the range from 0.85 to 0.95 times the flavor-specific
decay lifetime. With this allowed range a 2.9\% uncertainty on the efficiencies for $\Bsb$ decays to \CP 
eigenstates is found. The average $\Bs$ lifetime is known only to a precision of 3\%, which leads to a 1.5\% uncertainty 
on the selection efficiencies for $\Bs$ decays to flavor-specific final states. The $\Bz$ and $\Bm$
lifetimes are known with sufficient precision that the associated uncertainty is negligible.

Several of the efficiency factors are estimated from simulation.
Most, but not all, of the associated systematic uncertainties cancel due to the similar 
or identical final states for the signal and normalization modes. 
For modes with an unequal number of tracks in the final state, a 1\% uncertainty
due to small differences in the IP resolution between data and simulation is assigned.
The efficiency of the VS~\chisq requirement is checked using the large $\btodsd$
signal in data, and the agreement to within 1\% with the efficiency from simulation is the 
assigned uncertainty. For $\br(\btodzds)/\br(\btodsd)$, a 1\% uncertainty is attributed
to the efficiency of track reconstruction. For $\br(\bstodzdz)/\br(\btodzds)$,
the one fewer track in the $\Dz(K\pi)\Dzb(K\pi)$ final state is offset by the one extra track in
$\Dz(K\pi)\Dzb(K\pi\pi\pi)$, relative to $\Dz(K\pi)\Dsm(KK\pi)$, leading to a negligible tracking
uncertainty. The mass resolution in data is slightly larger than in simulation, resulting in slightly different
efficiencies for the reconstructed $\Dz$, $\Dp$ and $\Dsp$ invariant masses to lie within 25~\mevcc of their known 
masses. This introduces a maximum of 1\% uncertainty on the relative branching fractions.
To estimate the uncertainty on the trigger efficiencies determined from simulation, the hadron trigger 
efficiency ratios were also determined using data.
These efficiencies were measured using trigger-unbiased samples of kaons and pions identified in $\Dstarp\to\Dz\pip$ 
decays. Using this alternative procedure, we find that the simulated trigger efficiency ratios have an 
uncertainty of 2\%.
The combined systematic uncertainties in the efficiencies obtained from simulation are given 
in Table~\ref{tab:syst}.

The limited sizes of the $\B\to D\pim$ calibration samples lead to uncertainties in the BDT efficiencies.
The uncertainties on the ratios vary from 1.0\% to 2.0\%.
The uncertainty on the efficiency of the $D_{(s)}$ and $\Lc$ vetoes is dominated by the PID efficiencies,
but they only apply to the subset of $D$ candidates that fall within the mass window of two charm hadrons,
{\it e.g.,} both the $\Dp$ and $\Dsp$ mesons, which occurs about 20\% of the time for $\Dsp$ decays. 
Taking this fraction and the uncertainty in the PID efficiency into account, the veto efficiencies are 
estimated to have uncertainties of 1.0\% for the $\Dp$ veto, $0.5\%$ for the $\Dsp$ veto, and $0.3\%$ for the $\Lc$ veto. 

The fit model is validated using simulated experiments, and is found to be unbiased.
To assess the uncertainty due to the imperfect knowledge of the various parameters used in the fit model, 
a number of variations are investigated. The only non-negligible uncertainties
are due to the $B\to D\Km\Kp\pim$ background contribution, which is varied from 0\% to 2\%, and 
the cross-feed from ${\bstodsd}$ decays into the ${\bstodsds}$ sample. The uncertainty varies from 1.7\% to 2.1\%.
For $\br(\bstodd)/\br(\btodd)$ and $\br(\bstodsd)/\br(\btodsd)$, we assign an uncertainty of 0.5\%,
which accounts for potentially small differences in the signal shape for $\Bzb$
and $\Bsb$ decays (due to the $\Bz$-$\Bs$ mass difference). Lastly, the finite size of the
samples of simulated decays contributes 3\% uncertainty to all the measurements.
In total, the systematic uncertainties on 
the branching fraction ratios range from 5.5\% to 13.0\%, as indicated in Table~\ref{tab:syst}.

\begin{table*}[tb]
\begin{center}
\caption{\small{Sources of systematic uncertainty and their values (in \%) for the ratios of branching fractions of 
the indicated decays. For $\br(\Bzb_{(s)}\to\Dz\Dzb)/\br(\btodzds)$, the error on $f_d/f_s$ only applies to the
$\bstodzdz$ decay, as indicated by the values in parentheses.}}
%\vspace{5mm}
\begin{tabular}{lccccc}
\hline\hline
 & & & & & \\[-1.85ex]
Source          & $\frac{\bstodsds}{\btodsd}$ & $\frac{\bstodsd}{\btodsd}$ & $\frac{\bstodd}{\btodd}$  & $\frac{\Bzb_{(s)}\to\Dz\Dzb}{\btodzds}$ & $\frac{\btodzds}{\btodsd}$ \\
 & & & & & \\[-1.85ex]
\hline
$f_d/f_s$       &         5.7     &              5.7   &             5.7              &         $-$ (5.7)          &       $-$     \\
$\br(D)$         &         $-$      &             5.3   &             5.3              &         10.2         &       2.5     \\
$B$ meson lifetimes   &         2.9      &              1.5   &             2.9              &         2.9          &       $-$     \\
Eff. from simulation  & 2.4      &              $-$   &             $-$              &         2.2          &       2.6     \\
BDT selection         &         1.4      &              $-$   &             $-$              &         2.2          &       1.4     \\
Cross-feed vetoes      &         0.6      &              $-$   &             $-$              &         0.5          &       1.0     \\
$D$ mass resolution   &         1.0      &              $-$   &             $-$              &         1.0          &       1.0     \\
Fit model       &         2.1      &              0.5   &             0.5              &         1.7          &       2.1     \\
Simulated sample size  &         3.0      &              3.0   &             3.0              &         3.0          &       3.0     \\
\hline          
Total           &         8.0    &              8.5    &             8.9              &         11.7\,(13.0)         &       5.5     \\
\hline\hline
\end{tabular}
\label{tab:syst}
\end{center}
\end{table*}

\section{Discussion and summary}

First observations and measurements of the relative branching fractions for the decays ${\bstodd}$, 
${\bstodsd}$ and ${\bstodzdz}$ 
have been presented, along with measurements of $\br(\bstodsds)$ and
$\br(\btodzds)$. Taking the world average values for $\br(\btodsd)=(7.2\pm0.8)\times10^{-3}$~\cite{Beringer:1900zz}, 
the absolute branching fractions are
\begin{align*}
\br(\btodzds) &=  (8.6\pm0.2\,(\rstat)\pm0.4\,(\rsyst)\pm1.0\,({\rm norm}))\times10^{-3},  \nonumber \\
\br(\bstodsds) &= (4.0\pm0.2\,(\rstat)\pm0.3\,(\rsyst)\pm0.4\,({\rm norm}))\times10^{-3}.  \nonumber 
\end{align*}
\noindent  The third uncertainty reflects the precision of the branching fraction for the normalization
mode. These measurements are consistent with, and more precise than, both the current 
world average measurements~\cite{Beringer:1900zz} as well as the more recent measurement of 
${\br(\bstodsds)}$~\cite{Esen:2012yz}.

The measured value of $\br(\bstodsds)/\br(\btodsd)=0.55\pm0.06$ is significantly lower than the
naive expectation of unity for the case that both decays are dominated by tree amplitudes (see Fig.~\ref{fig:feyn}(a)),
assuming small non-factorizable effects and comparable magnitudes of the $B_{(s)}\to D^+_{(s)}$ form 
factors~\cite{Bailey:2012rr}. Unlike $\btodsd$,
the $\bstodsds$ decay receives a contribution from the $W$-exchange process (see Fig.~\ref{fig:feyn}(b)), 
suggesting that this amplitude may not be negligible. Interestingly, when comparing the ${\bstodsds}$ and 
${\btodd}$ decays, which have the same set of amplitudes, one finds ${|V_{cd}/V_{cs}|^2\cdot\br(\bstodsds)/\br(\btodd)\sim1}$.

Using ${\br(\btodd)=(2.11\pm0.31)\times10^{-4}}$ and ${\br(\btodzds)=}(10.0\pm1.7)\times10^{-3}$~\cite{Beringer:1900zz}, 
the following values for the branching fractions are obtained
\begin{align*}
\br(\bstodd)   &= (2.2\pm0.4\,(\rstat)\pm0.2\,(\rsyst)\pm0.3\,({\rm norm}))\times10^{-4},  \nonumber \\
\br(\bstodzdz) &= (1.9\pm0.3\,(\rstat)\pm0.3\,(\rsyst)\pm0.3\,({\rm norm}))\times10^{-4},  \nonumber \\
\br(\btodzdz)   &= (1.4\pm0.6\,(\rstat)\pm0.2\,(\rsyst)\pm0.2\,({\rm norm}))\times10^{-5}.  \nonumber 
\end{align*}
\noindent  
The first of these results disfavors the predicted values for $\br(\bstodd)$ in Refs.~\cite{Li:2003az,Eeg:2003yq}, which
are about 5--15 times larger than our measured value. The measured branching fractions are
about a factor of 2--3 larger than the predictions obtained by assuming that these decay amplitudes are 
dominated by rescattering~\cite{rosnerDsDs}. 
As discussed above for the $\br(\bstodsds)$ measurement, this may also suggest that the 
$W$-exchange amplitude contribution is not negligible in $B\to D\Dbar^{\prime}$ decays.
For precise quantitative comparisons of these
$\Bs$ branching fraction measurements to theoretical predictions, one should 
account for the different total widths of the \CP-even and \CP-odd final 
states~\cite{DeBruyn:2012wj}.

The Cabibbo suppressed $\bstodsd$ decay is also observed for the first time. Its absolute
branching fraction is
\begin{align*}
\br(\bstodsd)   &= (3.6\pm0.6\,(\rstat)\pm0.3\,(\rsyst)\pm0.4\,({\rm norm}))\times10^{-4}.  \nonumber 
\end{align*}
\noindent This value is consistent with the expected suppression of $|V_{cd}/V_{cs}|^2$.

The results reported here are based on an integrated luminosity of 1.0~$\ifb$. A data sample
with approximately 2.5 times larger yields in these modes has already been collected in 2012, and larger 
samples are anticipated in the next few years. These samples give good prospects for
\CP-violation measurements, 
lifetime studies, and obtaining a deeper understanding of the decay mechanisms that 
contribute to $b$-hadron decays.

\section*{Acknowledgements}

\noindent We express our gratitude to our colleagues in the CERN
accelerator departments for the excellent performance of the LHC. We
thank the technical and administrative staff at the LHCb
institutes. We acknowledge support from CERN and from the national
agencies: CAPES, CNPq, FAPERJ and FINEP (Brazil); NSFC (China);
CNRS/IN2P3 and Region Auvergne (France); BMBF, DFG, HGF and MPG
(Germany); SFI (Ireland); INFN (Italy); FOM and NWO (The Netherlands);
SCSR (Poland); ANCS/IFA (Romania); MinES, Rosatom, RFBR and NRC
``Kurchatov Institute'' (Russia); MinECo, XuntaGal and GENCAT (Spain);
SNSF and SER (Switzerland); NAS Ukraine (Ukraine); STFC (United
Kingdom); NSF (USA). We also acknowledge the support received from the
ERC under FP7. The Tier1 computing centres are supported by IN2P3
(France), KIT and BMBF (Germany), INFN (Italy), NWO and SURF (The
Netherlands), PIC (Spain), GridPP (United Kingdom). We are thankful
for the computing resources put at our disposal by Yandex LLC
(Russia), as well as to the communities behind the multiple open
source software packages that we depend on.

%\addcontentsline{toc}{section}{References}

\bibliographystyle{LHCb}
%\bibliography{main,LHCb-PAPER,LHCb-CONF}
\bibliography{main}

\ifx\mcitethebibliography\mciteundefinedmacro
\PackageError{LHCb.bst}{mciteplus.sty has not been loaded}
{This bibstyle requires the use of the mciteplus package.}\fi
\providecommand{\href}[2]{#2}
\begin{mcitethebibliography}{10}
\mciteSetBstSublistMode{n}
\mciteSetBstMaxWidthForm{subitem}{\alph{mcitesubitemcount})}
\mciteSetBstSublistLabelBeginEnd{\mcitemaxwidthsubitemform\space}
{\relax}{\relax}

\bibitem{Cabibbo:1963yz}
N.~Cabibbo, \ifthenelse{\boolean{articletitles}}{{\it {Unitary symmetry and
  leptonic decays}},
  }{}\href{http://dx.doi.org/10.1103/PhysRevLett.10.531}{Phys.\ Rev.\ Lett.\
  {\bf 10} (1963) 531}\relax
\mciteBstWouldAddEndPuncttrue
\mciteSetBstMidEndSepPunct{\mcitedefaultmidpunct}
{\mcitedefaultendpunct}{\mcitedefaultseppunct}\relax
\EndOfBibitem
\bibitem{Kobayashi:1973fv}
M.~Kobayashi and T.~Maskawa, \ifthenelse{\boolean{articletitles}}{{\it {CP
  violation in the renormalizable theory of weak interaction}},
  }{}\href{http://dx.doi.org/10.1143/PTP.49.652}{Prog.\ Theor.\ Phys.\  {\bf
  49} (1973) 652}\relax
\mciteBstWouldAddEndPuncttrue
\mciteSetBstMidEndSepPunct{\mcitedefaultmidpunct}
{\mcitedefaultendpunct}{\mcitedefaultseppunct}\relax
\EndOfBibitem
\bibitem{Aubert:2008ah}
BaBar collaboration, B.~Aubert {\em et~al.},
  \ifthenelse{\boolean{articletitles}}{{\it {Measurements of time-dependent CP
  asymmetries in $\Bz\to D^{(*)+} D^{(*)-}$ decays}},
  }{}\href{http://dx.doi.org/10.1103/PhysRevD.79.032002}{Phys.\ Rev.\  {\bf
  D79} (2009) 032002}, \href{http://arxiv.org/abs/0808.1866}{{\tt
  arXiv:0808.1866}}\relax
\mciteBstWouldAddEndPuncttrue
\mciteSetBstMidEndSepPunct{\mcitedefaultmidpunct}
{\mcitedefaultendpunct}{\mcitedefaultseppunct}\relax
\EndOfBibitem
\bibitem{Fratina:2007zk}
Belle collaboration, S.~Fratina {\em et~al.},
  \ifthenelse{\boolean{articletitles}}{{\it {Evidence for \CP violation in
  $\Bz\to\Dp\Dm$ decays}},
  }{}\href{http://dx.doi.org/10.1103/PhysRevLett.98.221802}{Phys.\ Rev.\ Lett.\
   {\bf 98} (2007) 221802}, \href{http://arxiv.org/abs/hep-ex/0702031}{{\tt
  arXiv:hep-ex/0702031}}\relax
\mciteBstWouldAddEndPuncttrue
\mciteSetBstMidEndSepPunct{\mcitedefaultmidpunct}
{\mcitedefaultendpunct}{\mcitedefaultseppunct}\relax
\EndOfBibitem
\bibitem{Aleksan:1993qk}
R.~Aleksan {\em et~al.}, \ifthenelse{\boolean{articletitles}}{{\it {The decay
  $B\to D\Dstarb + \Dstar\Dbar$ in the heavy quark limit and tests of CP
  violation}}, }{}\href{http://dx.doi.org/10.1016/0370-2693(93)91588-E}{Phys.\
  Lett.\  {\bf B317} (1993) 173}\relax
\mciteBstWouldAddEndPuncttrue
\mciteSetBstMidEndSepPunct{\mcitedefaultmidpunct}
{\mcitedefaultendpunct}{\mcitedefaultseppunct}\relax
\EndOfBibitem
\bibitem{Sanda:1996pm}
{A. I. Sanda and Z. Z. Xing}, \ifthenelse{\boolean{articletitles}}{{\it
  {Towards determining $\phi_1$ with $B\to\D^{(*)}\Dbar^{(*)}$}},
  }{}\href{http://dx.doi.org/10.1103/PhysRevD.56.341}{Phys.\ Rev.\  {\bf D56}
  (1997) 341}, \href{http://arxiv.org/abs/hep-ph/9702297}{{\tt
  arXiv:hep-ph/9702297}}\relax
\mciteBstWouldAddEndPuncttrue
\mciteSetBstMidEndSepPunct{\mcitedefaultmidpunct}
{\mcitedefaultendpunct}{\mcitedefaultseppunct}\relax
\EndOfBibitem
\bibitem{Xing:1998ca}
{Z. Z. Xing}, \ifthenelse{\boolean{articletitles}}{{\it {Measuring CP violation
  and testing factorization in $\B_d\to D^{*\pm}D^{*\mp}$ and $B_s\to
  D_s^{*\pm}D_s^{*\mp}$ decays}},
  }{}\href{http://dx.doi.org/10.1016/S0370-2693(98)01285-4}{Phys.\ Lett.\  {\bf
  B443} (1998) 365}, \href{http://arxiv.org/abs/hep-ph/9809496}{{\tt
  arXiv:hep-ph/9809496}}\relax
\mciteBstWouldAddEndPuncttrue
\mciteSetBstMidEndSepPunct{\mcitedefaultmidpunct}
{\mcitedefaultendpunct}{\mcitedefaultseppunct}\relax
\EndOfBibitem
\bibitem{Xing:1999yx}
{Z. Z. Xing}, \ifthenelse{\boolean{articletitles}}{{\it {CP violation in
  $B_d\to\Dp\Dm,~\Dstarp\Dm,~\Dp\Dstarm$, and $\Dstarp\Dstarm$ decays}},
  }{}\href{http://dx.doi.org/10.1103/PhysRevD.61.014010}{Phys.\ Rev.\  {\bf
  D61} (2000) 014010}, \href{http://arxiv.org/abs/hep-ph/9907455}{{\tt
  arXiv:hep-ph/9907455}}\relax
\mciteBstWouldAddEndPuncttrue
\mciteSetBstMidEndSepPunct{\mcitedefaultmidpunct}
{\mcitedefaultendpunct}{\mcitedefaultseppunct}\relax
\EndOfBibitem
\bibitem{Pham:1999fy}
{X. Y. Pham and Z. Z. Xing}, \ifthenelse{\boolean{articletitles}}{{\it {CP
  asymmetries in ${B_{d}\to D^{*+} D^{*-}}$ and ${B_{s}\to\Dss\Dssm}$ decays: P
  wave dilution, penguin and rescattering effects}},
  }{}\href{http://dx.doi.org/10.1016/S0370-2693(99)00611-5}{Phys.\ Lett.\  {\bf
  B458} (1999) 375}, \href{http://arxiv.org/abs/hep-ph/9904360}{{\tt
  arXiv:hep-ph/9904360}}\relax
\mciteBstWouldAddEndPuncttrue
\mciteSetBstMidEndSepPunct{\mcitedefaultmidpunct}
{\mcitedefaultendpunct}{\mcitedefaultseppunct}\relax
\EndOfBibitem
\bibitem{Datta:2003va}
A.~Datta and D.~London, \ifthenelse{\boolean{articletitles}}{{\it {Extracting
  $\gamma$ from $B_d^0(t)\to\D^{(*)+}D^{(*)-}$ and
  $B_d^0\to\D_s^{(*)+}D_s^{(*)-}$ decays}},
  }{}\href{http://dx.doi.org/10.1016/j.physletb.2003.12.048}{Phys.\ Lett.\
  {\bf B584} (2004) 81}, \href{http://arxiv.org/abs/hep-ph/0310252}{{\tt
  arXiv:hep-ph/0310252}}\relax
\mciteBstWouldAddEndPuncttrue
\mciteSetBstMidEndSepPunct{\mcitedefaultmidpunct}
{\mcitedefaultendpunct}{\mcitedefaultseppunct}\relax
\EndOfBibitem
\bibitem{Fleischer:2007zn}
R.~Fleischer, \ifthenelse{\boolean{articletitles}}{{\it {Exploring CP violation
  and penguin effects through $B_d^0\to\D^{+}D^{-}$ and
  $B_s^0\to\D_s^{+}D_s^{-}$}},
  }{}\href{http://dx.doi.org/10.1140/epjc/s10052-007-0341-4}{Eur.\ Phys.\ J.\
  {\bf C51} (2007) 849}, \href{http://arxiv.org/abs/0705.4421}{{\tt
  arXiv:0705.4421}}\relax
\mciteBstWouldAddEndPuncttrue
\mciteSetBstMidEndSepPunct{\mcitedefaultmidpunct}
{\mcitedefaultendpunct}{\mcitedefaultseppunct}\relax
\EndOfBibitem
\bibitem{DeBruyn:2012wj}
K.~De~Bruyn {\em et~al.}, \ifthenelse{\boolean{articletitles}}{{\it {Branching
  ratio measurements of $B_s$ decays}},
  }{}\href{http://dx.doi.org/10.1103/PhysRevD.86.014027}{Phys.\ Rev.\  {\bf
  D86} (2012) 014027}, \href{http://arxiv.org/abs/1204.1735}{{\tt
  arXiv:1204.1735}}\relax
\mciteBstWouldAddEndPuncttrue
\mciteSetBstMidEndSepPunct{\mcitedefaultmidpunct}
{\mcitedefaultendpunct}{\mcitedefaultseppunct}\relax
\EndOfBibitem
\bibitem{Fleischer:2011cw}
R.~Fleischer and R.~Knegjens, \ifthenelse{\boolean{articletitles}}{{\it
  {Effective lifetimes of $B_s$ decays and their constraints on the
  $B_s^0$-$\bar B_s^0$ mixing parameters}}, }{}Eur.\ Phys.\ J.\  {\bf C71}
  (2011) 1789, \href{http://arxiv.org/abs/1109.5115}{{\tt
  arXiv:1109.5115}}\relax
\mciteBstWouldAddEndPuncttrue
\mciteSetBstMidEndSepPunct{\mcitedefaultmidpunct}
{\mcitedefaultendpunct}{\mcitedefaultseppunct}\relax
\EndOfBibitem
\bibitem{hfag}
Y.~Amhis {\em et~al.}, \ifthenelse{\boolean{articletitles}}{{\it {Averages of
  b-hadron, c-hadron, and tau-lepton properties as of early 2012}},
  }{}\href{http://arxiv.org/abs/1207.1158}{{\tt arXiv:1207.1158}}, More
  information is available at
  \href{http://www.slac.stanford.edu/xorg/hfag/}{www.slac.stanford.edu/xorg/hf%
ag}\relax
\mciteBstWouldAddEndPuncttrue
\mciteSetBstMidEndSepPunct{\mcitedefaultmidpunct}
{\mcitedefaultendpunct}{\mcitedefaultseppunct}\relax
\EndOfBibitem
\bibitem{Aaij:2012ns}
LHCb collaboration, R.~Aaij {\em et~al.},
  \ifthenelse{\boolean{articletitles}}{{\it {Measurement of the effective
  $\Bs\rightarrow\Kp\Km$ lifetime}},
  }{}\href{http://dx.doi.org/10.1016/j.physletb.2012.08.033}{Phys.\ Lett.\
  {\bf B716} (2012) 393}, \href{http://arxiv.org/abs/1207.5993}{{\tt
  arXiv:1207.5993}}\relax
\mciteBstWouldAddEndPuncttrue
\mciteSetBstMidEndSepPunct{\mcitedefaultmidpunct}
{\mcitedefaultendpunct}{\mcitedefaultseppunct}\relax
\EndOfBibitem
\bibitem{Aaij:2012nt}
LHCb collaboration, R.~Aaij {\em et~al.},
  \ifthenelse{\boolean{articletitles}}{{\it {Measurement of the $\Bs$ effective
  lifetime in the $\jpsi f_0(980)$ final state}},
  }{}\href{http://dx.doi.org/10.1103/PhysRevLett.109.152002}{Phys.\ Rev.\
  Lett.\  {\bf 109} (2012) 152002}, \href{http://arxiv.org/abs/1207.0878}{{\tt
  arXiv:1207.0878}}\relax
\mciteBstWouldAddEndPuncttrue
\mciteSetBstMidEndSepPunct{\mcitedefaultmidpunct}
{\mcitedefaultendpunct}{\mcitedefaultseppunct}\relax
\EndOfBibitem
\bibitem{rosnerDsDs}
M.~Gronau, D.~London, and J.~Rosner, \ifthenelse{\boolean{articletitles}}{{\it
  {Rescattering contributions to rare B meson decays}},
  }{}\href{http://arxiv.org/abs/1211.5785}{{\tt arXiv:1211.5785}}\relax
\mciteBstWouldAddEndPuncttrue
\mciteSetBstMidEndSepPunct{\mcitedefaultmidpunct}
{\mcitedefaultendpunct}{\mcitedefaultseppunct}\relax
\EndOfBibitem
\bibitem{Beringer:1900zz}
Particle Data Group, J.~Beringer {\em et~al.},
  \ifthenelse{\boolean{articletitles}}{{\it {Review of particle physics
  (RPP)}}, }{}\href{http://dx.doi.org/10.1103/PhysRevD.86.010001}{Phys.\ Rev.\
  {\bf D86} (2012) 010001}\relax
\mciteBstWouldAddEndPuncttrue
\mciteSetBstMidEndSepPunct{\mcitedefaultmidpunct}
{\mcitedefaultendpunct}{\mcitedefaultseppunct}\relax
\EndOfBibitem
\bibitem{Aaij:2012as}
LHCb collaboration, R.~Aaij {\em et~al.},
  \ifthenelse{\boolean{articletitles}}{{\it {Measurement of $b$-hadron
  branching fractions for two-body decays into charmless charged hadrons}},
  }{}\href{http://dx.doi.org/10.1007/JHEP10(2012)037}{JHEP {\bf 10} (2012)
  037}, \href{http://arxiv.org/abs/1206.2794}{{\tt arXiv:1206.2794}}\relax
\mciteBstWouldAddEndPuncttrue
\mciteSetBstMidEndSepPunct{\mcitedefaultmidpunct}
{\mcitedefaultendpunct}{\mcitedefaultseppunct}\relax
\EndOfBibitem
\bibitem{Li:2003az}
Y.~Li, C.-D. Lu, and Z.-J. Xiao, \ifthenelse{\boolean{articletitles}}{{\it
  {Rare decays $\Bz\to D_s^{(*)+}D_s^{(*)-}$ and $\Bs\to D^{(*)+}D^{(*)-}$ in
  perturbative QCD approach}},
  }{}\href{http://dx.doi.org/10.1088/0954-3899/31/3/007}{J.\ Phys.\  {\bf G31}
  (2005) 273}, \href{http://arxiv.org/abs/hep-ph/0308243}{{\tt
  arXiv:hep-ph/0308243}}\relax
\mciteBstWouldAddEndPuncttrue
\mciteSetBstMidEndSepPunct{\mcitedefaultmidpunct}
{\mcitedefaultendpunct}{\mcitedefaultseppunct}\relax
\EndOfBibitem
\bibitem{Eeg:2003yq}
J.~Eeg, S.~Fajfer, and A.~Hiorth, \ifthenelse{\boolean{articletitles}}{{\it {On
  the color suppressed decay modes $\Bzb\to\Dsp\Dsm$ and $\Bsb\to\Dp\Dm$}},
  }{}\href{http://dx.doi.org/10.1016/j.physletb.2003.07.013}{Phys.\ Lett.\
  {\bf B570} (2003) 46}, \href{http://arxiv.org/abs/hep-ph/0304112}{{\tt
  arXiv:hep-ph/0304112}}\relax
\mciteBstWouldAddEndPuncttrue
\mciteSetBstMidEndSepPunct{\mcitedefaultmidpunct}
{\mcitedefaultendpunct}{\mcitedefaultseppunct}\relax
\EndOfBibitem
\bibitem{Alves:2008zz}
LHCb collaboration, A.~A. Alves~Jr. {\em et~al.},
  \ifthenelse{\boolean{articletitles}}{{\it {The \lhcb detector at the LHC}},
  }{}\href{http://dx.doi.org/10.1088/1748-0221/3/08/S08005}{JINST {\bf 3}
  (2008) S08005}\relax
\mciteBstWouldAddEndPuncttrue
\mciteSetBstMidEndSepPunct{\mcitedefaultmidpunct}
{\mcitedefaultendpunct}{\mcitedefaultseppunct}\relax
\EndOfBibitem
\bibitem{LHCbRich}
M.~Adinolfi {\em et~al.}, \ifthenelse{\boolean{articletitles}}{{\it
  {Performance of the LHCb RICH detector at the LHC}},
  }{}\href{http://arxiv.org/abs/1211.6759}{{\tt arXiv:1211.6759}}, (submitted
  to Eur.~Phys.~J.~{\bf C})\relax
\mciteBstWouldAddEndPuncttrue
\mciteSetBstMidEndSepPunct{\mcitedefaultmidpunct}
{\mcitedefaultendpunct}{\mcitedefaultseppunct}\relax
\EndOfBibitem
\bibitem{Aaij:2012me}
R.~Aaij {\em et~al.}, \ifthenelse{\boolean{articletitles}}{{\it {The \lhcb
  trigger and its performance}}, }{}\href{http://arxiv.org/abs/1211.3055}{{\tt
  arXiv:1211.3055}}, (submitted to JINST)\relax
\mciteBstWouldAddEndPuncttrue
\mciteSetBstMidEndSepPunct{\mcitedefaultmidpunct}
{\mcitedefaultendpunct}{\mcitedefaultseppunct}\relax
\EndOfBibitem
\bibitem{bbdt}
V.~V. Gligorov and M.~Williams, \ifthenelse{\boolean{articletitles}}{{\it
  {Efficient, reliable and fast high-level triggering using a bonsai boosted
  decision tree}}, }{}\href{http://arxiv.org/abs/1210.6861}{{\tt
  arXiv:1210.6861}}, (submitted to JINST)\relax
\mciteBstWouldAddEndPuncttrue
\mciteSetBstMidEndSepPunct{\mcitedefaultmidpunct}
{\mcitedefaultendpunct}{\mcitedefaultseppunct}\relax
\EndOfBibitem
\bibitem{Sjostrand:2006za}
T.~Sj\"{o}strand, S.~Mrenna, and P.~Z. Skands,
  \ifthenelse{\boolean{articletitles}}{{\it {PYTHIA 6.4 Physics and manual}},
  }{}\href{http://dx.doi.org/10.1088/1126-6708/2006/05/026}{JHEP {\bf 05}
  (2006) 026}, \href{http://arxiv.org/abs/hep-ph/0603175}{{\tt
  arXiv:hep-ph/0603175}}\relax
\mciteBstWouldAddEndPuncttrue
\mciteSetBstMidEndSepPunct{\mcitedefaultmidpunct}
{\mcitedefaultendpunct}{\mcitedefaultseppunct}\relax
\EndOfBibitem
\bibitem{LHCb-PROC-2010-056}
I.~Belyaev {\em et~al.}, \ifthenelse{\boolean{articletitles}}{{\it {Handling of
  the generation of primary events in \gauss, the \lhcb simulation framework}},
  }{}\href{http://dx.doi.org/10.1109/NSSMIC.2010.5873949}{Nuclear Science
  Symposium Conference Record (NSS/MIC) {\bf IEEE} (2010) 1155}\relax
\mciteBstWouldAddEndPuncttrue
\mciteSetBstMidEndSepPunct{\mcitedefaultmidpunct}
{\mcitedefaultendpunct}{\mcitedefaultseppunct}\relax
\EndOfBibitem
\bibitem{Lange:2001uf}
D.~J. Lange, \ifthenelse{\boolean{articletitles}}{{\it {The EvtGen particle
  decay simulation package}},
  }{}\href{http://dx.doi.org/10.1016/S0168-9002(01)00089-4}{Nucl.\ Instrum.\
  Meth.\  {\bf A462} (2001) 152}\relax
\mciteBstWouldAddEndPuncttrue
\mciteSetBstMidEndSepPunct{\mcitedefaultmidpunct}
{\mcitedefaultendpunct}{\mcitedefaultseppunct}\relax
\EndOfBibitem
\bibitem{Golonka:2005pn}
P.~Golonka and Z.~Was, \ifthenelse{\boolean{articletitles}}{{\it {PHOTOS Monte
  Carlo: a precision tool for QED corrections in $Z$ and $W$ decays}},
  }{}\href{http://dx.doi.org/10.1140/epjc/s2005-02396-4}{Eur.\ Phys.\ J.\  {\bf
  C45} (2006) 97}, \href{http://arxiv.org/abs/hep-ph/0506026}{{\tt
  arXiv:hep-ph/0506026}}\relax
\mciteBstWouldAddEndPuncttrue
\mciteSetBstMidEndSepPunct{\mcitedefaultmidpunct}
{\mcitedefaultendpunct}{\mcitedefaultseppunct}\relax
\EndOfBibitem
\bibitem{Allison:2006ve}
GEANT4 collaboration, J.~Allison {\em et~al.},
  \ifthenelse{\boolean{articletitles}}{{\it {Geant4 developments and
  applications}}, }{}\href{http://dx.doi.org/10.1109/TNS.2006.869826}{IEEE
  Trans.\ Nucl.\ Sci.\  {\bf 53} (2006) 270}\relax
\mciteBstWouldAddEndPuncttrue
\mciteSetBstMidEndSepPunct{\mcitedefaultmidpunct}
{\mcitedefaultendpunct}{\mcitedefaultseppunct}\relax
\EndOfBibitem
\bibitem{Agostinelli:2002hh}
GEANT4 collaboration, S.~Agostinelli {\em et~al.},
  \ifthenelse{\boolean{articletitles}}{{\it {GEANT4: A simulation toolkit}},
  }{}\href{http://dx.doi.org/10.1016/S0168-9002(03)01368-8}{Nucl.\ Instrum.\
  Meth.\  {\bf A506} (2003) 250}\relax
\mciteBstWouldAddEndPuncttrue
\mciteSetBstMidEndSepPunct{\mcitedefaultmidpunct}
{\mcitedefaultendpunct}{\mcitedefaultseppunct}\relax
\EndOfBibitem
\bibitem{LHCb-PROC-2011-006}
M.~Clemencic {\em et~al.}, \ifthenelse{\boolean{articletitles}}{{\it {The \lhcb
  simulation application, \gauss: design, evolution and experience}},
  }{}\href{http://dx.doi.org/10.1088/1742-6596/331/3/032023}{{J.\ of Phys.\ :
  Conf.\ Ser.\ } {\bf 331} (2011) 032023}\relax
\mciteBstWouldAddEndPuncttrue
\mciteSetBstMidEndSepPunct{\mcitedefaultmidpunct}
{\mcitedefaultendpunct}{\mcitedefaultseppunct}\relax
\EndOfBibitem
\bibitem{Narsky:2005hn}
I.~Narsky, \ifthenelse{\boolean{articletitles}}{{\it {Optimization of signal
  significance by bagging decision trees}},
  }{}\href{http://arxiv.org/abs/physics/0507157}{{\tt
  arXiv:physics/0507157}}\relax
\mciteBstWouldAddEndPuncttrue
\mciteSetBstMidEndSepPunct{\mcitedefaultmidpunct}
{\mcitedefaultendpunct}{\mcitedefaultseppunct}\relax
\EndOfBibitem
\bibitem{Narsky:2005hn2}
I.~Narsky, \ifthenelse{\boolean{articletitles}}{{\it {StatPatternRecognition: a
  C++ package for statistical analysis of high energy physics data}},
  }{}\href{http://arxiv.org/abs/physics/0507143}{{\tt
  arXiv:physics/0507143}}\relax
\mciteBstWouldAddEndPuncttrue
\mciteSetBstMidEndSepPunct{\mcitedefaultmidpunct}
{\mcitedefaultendpunct}{\mcitedefaultseppunct}\relax
\EndOfBibitem
\bibitem{Pivk:2004ty}
M.~Pivk and F.~R. Le~Diberder, \ifthenelse{\boolean{articletitles}}{{\it
  {sPlot: a statistical tool to unfold data distributions}},
  }{}\href{http://dx.doi.org/10.1016/j.nima.2005.08.106}{Nucl.\ Instrum.\
  Meth.\  {\bf A555} (2005) 356},
  \href{http://arxiv.org/abs/physics/0402083}{{\tt
  arXiv:physics/0402083}}\relax
\mciteBstWouldAddEndPuncttrue
\mciteSetBstMidEndSepPunct{\mcitedefaultmidpunct}
{\mcitedefaultendpunct}{\mcitedefaultseppunct}\relax
\EndOfBibitem
\bibitem{Skwarnicki:1986xj}
T.~Skwarnicki, {\em {A study of the radiative cascade transitions between the
  Upsilon-prime and Upsilon resonances}}, PhD thesis, Institute of Nuclear
  Physics, Krakow, 1986,
  {\href{http://inspirehep.net/record/230779/files/230779.pdf}{DESY-F31-86-02}%
}\relax
\mciteBstWouldAddEndPuncttrue
\mciteSetBstMidEndSepPunct{\mcitedefaultmidpunct}
{\mcitedefaultendpunct}{\mcitedefaultseppunct}\relax
\EndOfBibitem
\bibitem{Drutskoy:2002ib}
Belle Collaboration, A.~Drutskoy {\em et~al.},
  \ifthenelse{\boolean{articletitles}}{{\it {Observation of $B\to D^{(*)}\Km
  K^{0(*)}$ decays}},
  }{}\href{http://dx.doi.org/10.1016/S0370-2693(02)02373-0}{Phys.\ Lett.\  {\bf
  B542} (2002) 171}, \href{http://arxiv.org/abs/hep-ex/0207041}{{\tt
  arXiv:hep-ex/0207041}}\relax
\mciteBstWouldAddEndPuncttrue
\mciteSetBstMidEndSepPunct{\mcitedefaultmidpunct}
{\mcitedefaultendpunct}{\mcitedefaultseppunct}\relax
\EndOfBibitem
\bibitem{LHCb-PAPER-2012-037}
LHCb collaboration, R.~Aaij {\em et~al.},
  \ifthenelse{\boolean{articletitles}}{{\it {Measurement of the ratio of
  fragmentation fractions $f_s/f_d$ and dependence on $B$ meson kinematics}},
  }{}\href{http://arxiv.org/abs/1301.5286}{{\tt arXiv:1301.5286}}, (submitted
  to JHEP)\relax
\mciteBstWouldAddEndPuncttrue
\mciteSetBstMidEndSepPunct{\mcitedefaultmidpunct}
{\mcitedefaultendpunct}{\mcitedefaultseppunct}\relax
\EndOfBibitem
\bibitem{Lenz:2011ti}
A.~Lenz and U.~Nierste, \ifthenelse{\boolean{articletitles}}{{\it {Numerical
  updates of lifetimes and mixing parameters of B mesons}},
  }{}\href{http://arxiv.org/abs/1102.4274}{{\tt arXiv:1102.4274}}, proceedings
  of the $6^{th}$ International Workshop in the CKM Unitarity Triangle,
  Warwick, U.K., Sept. 6-10, 2010\relax
\mciteBstWouldAddEndPuncttrue
\mciteSetBstMidEndSepPunct{\mcitedefaultmidpunct}
{\mcitedefaultendpunct}{\mcitedefaultseppunct}\relax
\EndOfBibitem
\bibitem{LHCb-CONF-2012-002}
LHCb collaboration, \ifthenelse{\boolean{articletitles}}{{\it Tagged
  time-dependent angular analysis of ${\Bs\to\jpsi\phi}$ decays at \lhcb}, }{}
  \href{http://cdsweb.cern.ch/search?p=LHCb-CONF-2012-002&f=reportnumber&actio%
n_search=Search&c=LHCb+Reports&c=LHCb+Conference+Proceedings&c=LHCb+Conference%
+Contributions&c=LHCb+Notes&c=LHCb+Theses&c=LHCb+Papers}
  {LHCb-CONF-2012-002}\relax
\mciteBstWouldAddEndPuncttrue
\mciteSetBstMidEndSepPunct{\mcitedefaultmidpunct}
{\mcitedefaultendpunct}{\mcitedefaultseppunct}\relax
\EndOfBibitem
\bibitem{Esen:2012yz}
Belle collaboration, S.~Esen {\em et~al.},
  \ifthenelse{\boolean{articletitles}}{{\it {Precise measurement of the
  branching fractions for $B_s\to D_s^{(*)+} D_s^{(*)-}$ and first measurement
  of the $D_s^{*+} D_s^{*-}$ polarization using $e^+e^-$ collisions}},
  }{}\href{http://arxiv.org/abs/1208.0323}{{\tt arXiv:1208.0323}}\relax
\mciteBstWouldAddEndPuncttrue
\mciteSetBstMidEndSepPunct{\mcitedefaultmidpunct}
{\mcitedefaultendpunct}{\mcitedefaultseppunct}\relax
\EndOfBibitem
\bibitem{Bailey:2012rr}
J.~A. Bailey {\em et~al.}, \ifthenelse{\boolean{articletitles}}{{\it {$B_s\to
  D_s/B\to D$ semileptonic form-factor ratios and their application to
  $\br(B^0_s\to \mu^+\mu^-)$}},
  }{}\href{http://dx.doi.org/10.1103/PhysRevD.85.114502,
  10.1103/PhysRevD.86.039904}{Phys.\ Rev.\  {\bf D85} (2012) 114502},
  \href{http://arxiv.org/abs/1202.6346}{{\tt arXiv:1202.6346}}\relax
\mciteBstWouldAddEndPuncttrue
\mciteSetBstMidEndSepPunct{\mcitedefaultmidpunct}
{\mcitedefaultendpunct}{\mcitedefaultseppunct}\relax
\EndOfBibitem
\end{mcitethebibliography}

\end{document}